\documentclass[aps,showpacs]{revtex4}
\usepackage{amsmath,amsfonts,bm}
\usepackage[dvips]{graphicx}

\begin{document}

\title{Loop series for discrete statistical models on graphs}

\author{Michael Chertkov $^{(a)}$ and Vladimir Y. Chernyak $^{(b)}$}

\affiliation{$^{(a)}$ Theoretical Division and Center for Nonlinear
Studies,
LANL,  Los Alamos, NM 87545\\
$^{(b)}$ Department of Chemistry, Wayne State University, 5101 Cass
Ave, Detroit, MI 48202\\
E-mails: chertkov@lanl.gov,
chernyak@chem.wayne.edu}

\date{\today}

\begin{abstract}
In this paper we present the derivation details, logic, and
motivation for the loop calculus introduced in \cite{06CCa}.
Generating functions for three inter-related discrete statistical
models are expressed in terms of a finite series. The first term in
the series corresponds to the Bethe-Peierls/Belief-Propagation (BP)
contribution, the other terms are labeled by loops on the factor
graph. All loop contributions are simple rational functions of spin
correlation functions calculated within the BP approach. We discuss
two alternative derivations of the loop series. One approach
implements a set of local auxiliary integrations over continuous
fields with the BP contribution corresponding to an integrand
saddle-point value. The integrals are replaced by sums in the
complimentary approach, briefly explained in \cite{06CCa}. Local
gauge symmetry transformations that clarify an important invariant
feature of the BP solution, are revealed in both approaches. The
individual terms change under the gauge transformation while the
partition function remains invariant. The requirement for all
individual terms to be non-zero only for closed loops in the factor
graph (as opposed to paths with loose ends) is equivalent to fixing
the first term in the series to be exactly equal to the BP
contribution. Further applications of the loop calculus to problems
in statistical physics, computer and information sciences are
discussed.
\end{abstract}

\pacs{05.50.+q,89.70.+C}



\maketitle

One practically useful yet generally heuristic approach used for
calculations of observables (correlation functions) in discrete
statistical physics models, e.g. Ising model, is related to the
so-called Bethe-Peierls (BP) approximation \cite{35Bet,36Pei,82Bax}.
The BP approach is exact for graphs that do not contain loops,
usually referred to as trees; otherwise the approach is approximate.
The ad-hoc approach can also be re-stated in a variational form
\cite{51Kik,91Mor,05YFW}. A similar tree-based method in information
science has been developed by Gallager \cite{63Gal,68Gal} in the
context of error-correction theory. Gallager introduced the
so-called Low-Density-Parity-Check (LDPC) codes, defined on locally
tree-like Tanner graphs. The problem of ideal decoding, i.e.
restoring the most probable pre-image out of the exponentially large
pool of candidates, is identical to solving a statistical model on
the graph \cite{89Sou}. An approximate yet efficient
Belief-Propagation  decoding algorithm introduced by Gallager
constitutes an iterative solution of the Bethe-Peierls equations
derived as if the statistical problem was defined on a tree that
locally represents the Tanner graph. We utilize this abbreviation
coincidence to call Bethe-Peierls and Belief-Propagation equations
by the same acronym -- BP. Recent resurgence of interest to LDPC
codes \cite{99Mac,03RU},  as well as proliferation of the BP
approach to other areas of information and computer science, e.g.
artificial intelligence \cite{88Pea} and combinatorial optimization
\cite{02MPZ,02MZ,04BZ}, where interesting statistical models on
graphs with long loops appear, made the BP approach to be one of the
most interesting and hot research topics in modern information and
computer sciences.

In spite of the lack of analytical control in the general case of graphs/models with loops, the BP
approximation and the corresponding algorithm provide remarkably accurate results. Based on this
empirical observation one would expect an existence of a hidden mathematical structure that can
rationalize an inessential, sub-leading role of the corrections associated with the loops. Besides,
an in depth understanding of the BP success would also provide a practical guidance for improving
BP even further by accounting for non-local loop-related correlations. However, with the exception
of two recent papers \cite{05MR,05PS}, the discussion of this important point was largely
superficial and anecdotal. The Ising model (pair-wise interactions between the bits) on a graph
with loops has been considered by Montanari and Rizzo \cite{05MR}, where a set of exact equations
has been derived that relates the correlation functions to each other. This system of equations is
under-defined; however, if irreducible correlations are neglected the BP result is restored. This
feature has been used \cite{05MR} to generate a perturbative expansion for corrections to BP in
terms of irreducible correlations. A complementary approach for the Ising model on a lattice has
been taken by Parisi and Slanina \cite{05PS}, who utilized an integral representation developed by
Efetov \cite{90Efe} in early nineties. The saddle-point for the integral representation used in
\cite{05PS} turns out to be exactly the BP solution. Calculating perturbative corrections to
magnetization, the authors of \cite{05PS} encountered divergences in their representation for the
partition function, however, the divergences canceled out from the leading order correction to the
magnetization revealing a sensible loop correction to BP. These papers, \cite{05MR} and
\cite{05PS}, became important initial steps towards calculating and understanding loop corrections
to BP. However, both approaches are very far from being complete and problem-free. Thus,
\cite{05MR} lacks an invariant representation in terms of the partition function. Instead it
requires operating with correlation functions. Besides, the complexity of the equations related to
the higher-order corrections rapidly grows with the order. The complimentary approach of
\cite{05PS} contains dangerous, since lacking analytical control, divergences (zero modes), which
constitutes a very problematic symptom for any field theory. Both \cite{05MR} and \cite{05PS} focus
on the Ising pair-wise interaction model. The extensions of the proposed methods to the most
interesting from the information theory viewpoint multi-bit interaction cases do not look
straightforward. Finally, the approaches of \cite{05MR} and \cite{05PS}, if extended to
higher-order corrections, will result in infinite series. Re-summing the corrections in all orders,
so that the result is presented in terms of a finite series, does not look feasible within the
proposed techniques.

In \cite{06CCa} we suggested an ultimate way to account for loop
corrections to BP. We represented the partition function for a
general discrete statistical model defined on a finite factor graph
in terms of a series decomposition. The most remarkable feature of
the suggested exact decomposition, that does not appear within the
previous approaches \cite{05MR,05PS}, is the representation of the
partition function as a {\it finite} (!!) series with the first term
being exactly represented by the BP solution. All higher-order terms
are labeled by generalized loops in the factor graph. A generalized
loop is defined as a possibly branching undirected path in the
factor graph that has no loose ends. Each term in the series is
represented as a product of local contributions along the loop, each
contribution being expressed explicitly in terms of some correlation
functions calculated within the BP approximation.

The present manuscript generalizes and details the approach of
\cite{06CCa}. In addition to explaining all technical details of the
loop series derivation of \cite{06CCa} we also provide an
alternative approach based on an integral representation for the
partition function. For the integral representation BP appears as a
result of applying the saddle-point approximation. We pay special
attention to clarifying the relation between the saddle-point
approximation for the integral and the Bethe Free energy approach of
\cite{05YFW}, as well as between the analysis of the Gaussian
corrections and the saddle-point. We also provide a technical
rationale for a formal gauge transformation in the integral
representation for the model partition function (transformation of
variables and decomposition of the integrand in a series) that
results in the loop series expression.

The integral representation approach is formulated for a bipartite
factor graph model \cite{01KFL,04Loe,05YFW} which is a particular
case of the general vertex model of \cite{01For} also considered in
the manuscript. For the presentation clarity we introduce a
bipartite vertex model, orientable vertex model, that is less
general than the general vertex model, yet constitutes a
generalization of the bipartite factor graph model \cite{Models}.
The vertex models are more general compared to the bipartite factor
graph model and allow a simpler derivation of the loop series using
an auxiliary discrete transformation (discrete Fourier transforms)
in place of its integral counterpart. We actually start the
technical part of the paper by describing a simpler and more compact
discrete variable representation before turning to a lengthy, still
ideologically important, integral (continuous variables)
counterpart.

The auxiliary degrees of freedoms, one per the graph edge, introduced within both
integral/continious and sum/discreet approaches, possess a gauge symmetry that allows an invariant
definition of the BP equation. Gauge transformation corresponding to the symmetry keeps the full
expression for the partition function invariant while changing the individual term of the series.
An individual term corresponds to a path on the graph that may generally contain some number of
loose ends. The BP equations can be viewed as conditions for fixing a special gauge that requires
all allowed paths (i.e. those who contribute to the series) to be nothing but generalized loops
that do not contain any loose ends. For this special gauge the BP approximation is described by the
first bare term in the loop series.

The formulation of BP as a gauge fixing condition also allows a
clear physical interpretation of the entire approach. Indeed, the
first bare term of the loop series can be viewed as a ``ground
state" that minimizes the Bethe free energy with loop corrections
being related to certain excited states described as along-the-loops
spin flips with respect to the ground state. Such interpretation of
the loop series makes our approach similar to the so-called
high-temperature expansion, where individual contributions
(diagrams) also correspond to close loops on the factor graph. There
is, however, a very important key difference between the loop series
and the high-temperature expansion. While the high-temperature
expansion starts with a trivial bare term (just unity in the
expansion of the partition function) the bare term in the loop
series is highly nontrivial. It is represented by the BP
approximation that already accounts for some local correlations in
the model.

The manuscript is organized as follows. In Section \ref{sec:Models}
we introduce our three basic models: bipartite factor-graph model,
orientable (bipartite) vertex, and general vertex models. Vertex
models are convenient generalizations of the bipartite factor-graph
model. In Section \ref{sec:Models} we also state our major result --
exact expressions for the models' partition functions in terms of
finite series, coined loop-series, over closed paths defined in the
models' graphs. The rest of the paper is devoted to derivations and
discussions of these results. Straightforward and simple derivation
of the loop series for the vertex model is described in Section
\ref{sec:Ver}. BP-equations emerge as a result of a requirement for
the finite series representation for the partition function of the
model to have the loop-series form (with no terms, correspondent to
a path with loose ends, present). In Section \ref{sec:Fac} we derive
the loop-series for the factor-graph model via an integral
representation. This derivation is more involved, however we present
it here in full as it allows to establish a relation between the
loop-calculus and other approaches in theoretical physics, e.g.
saddle point analysis. Section \ref{sec:Fac} contains a number of
Subsections. Integral representation for the partition function of
the factor-node model is introduced in Subsection \ref{subsec:Int}.
Subsection \ref{subsec:Bethe} describes the relation of the integral
representation to the Bethe free energy variational approach of
\cite{05YFW}. The latter is also briefly sketched in Appendix
\ref{app:Bethe}. In Subsection \ref{subsec:SP} we present an
approximate saddle-point analysis of the integral representation for
the partition function of the factor-graph model. Here we show that
the saddle-point is described by the BP equations. The Gaussian
approximation around the BP saddle-point is discussed in Subsection
\ref{subsec:Gauss}. Finally, the derivation of the exact loop-series
via the integral representation is described in Subsection
\ref{subsec:LoopInt}. Section \ref{sec:Con} is devoted to
conclusions where we also discuss possible generalizations as well
as practical utility of the loop-series/caclulus in
information/computer science and statistical physics. Appendix
\ref{app:single} illustrates the loop calculus on a simple example
of two bits, two checks bipartite factor graph model with single
loop.

\section{Loop series for the factor-graph and vertex models}
\label{sec:Models}

\subsection{Bipartite factor-graph model}
\label{subsec:Fac}

\begin{figure} [b]
\includegraphics[width=0.45\textwidth]{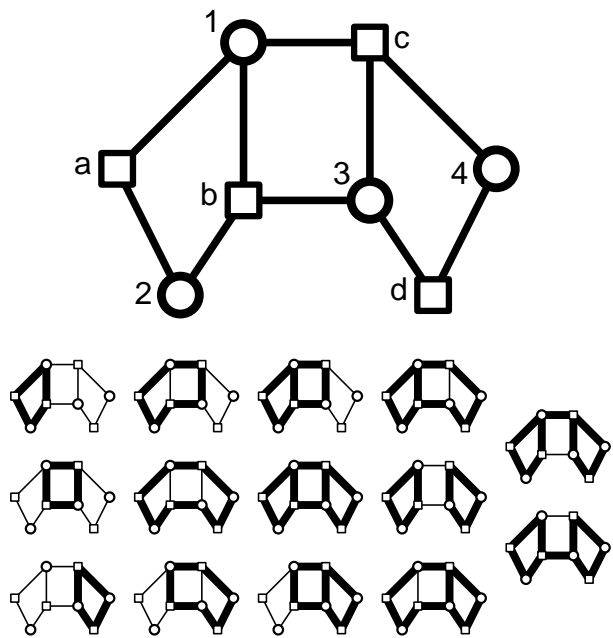}
\caption{Example of a factor graph. Fourteen possible marked paths
(generalized loops) for the example are shown in bold on the
bottom.} \label{paths}
\end{figure}

Consider a generic discrete statistical model, with configurations characterized by a set of binary
variables: $\sigma_i=\pm 1$, $i=1,\cdots,n$, which is factorized so that the probability
$p\{\sigma_i\}$ to find the system in the state $\{\sigma_i\}$ and the partition function $Z$ are
\begin{eqnarray}
p\{\sigma_i\}=Z^{-1}\prod\limits_\alpha
f_\alpha(\sigma_\alpha),\quad
Z=\sum\limits_{\{\sigma_i\}}\prod\limits_\alpha
f_\alpha(\sigma_\alpha),
 \label{p1}
\end{eqnarray}
where $\alpha$ labels non-negative and finite factor-functions
$f_\alpha$ with $\alpha=1,\ldots,m$ and $\sigma_\alpha$ represents a
subset of $\sigma_i$ variables. Relations between factor functions
(checks) and elementary discrete variables (bits), expressed as
$i\in\alpha$ and $\alpha\ni i$, can be conveniently represented in
terms of the system-specific factor graph. If $i\in\alpha$ we say
that the bit and the check are neighbors. An example of a factor
graph with $m=4$ that corresponds to
$p(\sigma_1,\sigma_2,\sigma_3,\sigma_4)=Z^{-1}f_a({\bm
\sigma}_a)f_b({\bm \sigma}_b)f_c({\bm \sigma}_c)f_d({\bm
\sigma}_d)$, where $\sigma_a\equiv\left(\sigma_1,\sigma_2\right)$,
$\sigma_b\equiv\left(\sigma_1,\sigma_2,\sigma_3\right)$,
$\sigma_c\equiv\left(\sigma_1,\sigma_3,\sigma_4\right)$,
$\sigma_d\equiv\left(\sigma_3,\sigma_4\right)$ and $\alpha=a,b,c,d$,
is shown in Fig.~\ref{paths}.
Any spin correlation function can be calculated using the partition function, $Z$, defined by
Eq.~(\ref{p1}). For example, the bit $i$ magnetization is expressed as
\begin{eqnarray}
\langle\sigma_i\rangle=\left.\frac{\partial \ln Z}{\partial
h_i}\right|_{h\to 0},
\end{eqnarray}
where the following transformation of a factor node function associated with a check $\alpha$
neighboring the bit $i$ is assumed: $f_\alpha(\sigma_\alpha)\to f_\alpha(\sigma_\alpha)
\exp(h_i\sigma_i)$.

\subsection{Vertex models}
\label{subsec:Ver}

In this subsection we discuss vertex models of two types,
orientable/bipartite and general. Similar to the factor graph model,
the vertex models are formulated in terms of Ising spin variables,
$\sigma=\pm$. However, while in the factor graph model spins reside
in the bit nodes, spins in the vertex models are assigned to the
edges. The orientable vertex model generalizes the factor-graph
model described in the previous Subsection, while the general vertex
model generalizes the orientable vertex model.

\subsubsection{Orientable Vertex model}
\label{subsubsec:Orient}

A graph is orientable if the whole family of its nodes, $X$, can be
partitioned in two sub-families, such that nodes of one sub-family
neighbors only nodes from the opposite sub-family. Also if a
connected graph is orientable, there are exactly two different
global orientations: A global orientation is chosen by picking some
node on a graph and identifying it as left (or right). Choosing an
orientation on an orientable graph partitions the set of nodes
$X=X_{L}\cup X_{R}$ into the subsets of left and right nodes,
referred to as bit nodes and check nodes, respectively. Ising
variables in the vertex model reside in the graph edges, i.e. the
configurations are defined by sets of Ising variables
$\sigma_{c}=\pm 1$ for $c\in X_{1}$. For a graph with a chosen
orientation it is also convenient to represent these variables as
$\sigma_{j\alpha}$, with $\alpha\in X_{R}$ and $j\in X_{L}$
representing the check (right) and bit (left) end of an edge. The
weight (probability) of a configuration is given by a product of
weights related to the nodes:
\begin{eqnarray}
 \label{weight-general} p_{\it ov}({\bm\sigma})=Z^{-1}_{\it ov}\prod_{j\in
 X_{L}}f_{j}\left({\bm\sigma}_{j}\right)\prod_{\alpha\in
 X_{R}}f_{\alpha}\left({\bm\sigma}_{\alpha}\right),\quad
 Z_{\it ov}=\sum\limits_{\{\bm \sigma\}} \prod_{j\in
 X_{L}}f_{j}\left({\bm\sigma}_{j}\right)\prod_{\alpha\in
 X_{R}}f_{\alpha}\left({\bm\sigma}_{\alpha}\right).
\end{eqnarray}

A particular example of the oriented vertex model defined for the graph shown in Fig.~\ref{paths}
corresponds to $p\sim f_1(\sigma_{1a},\sigma_{1b},\sigma_{1c})f_2(\sigma_{2a},\sigma_{2b})
f_3(\sigma_{3b},\sigma_{3c},\sigma_{3d})f_4(\sigma_{4c},\sigma_{4d})
f_a(\sigma_{a1},\sigma_{a2})f_b(\sigma_{b1},\sigma_{b2},\sigma_{b3})
f_c(\sigma_{c1},\sigma_{c3},\sigma_{c4})f_d(\sigma_{d3},\sigma_{d4})$, where we do not
differentiate between the bits and checks and the index order  for a spin defined on the graph edge
is not important.

Obviously, the oriented vertex model (\ref{weight-general}) turns
into the factor-graph model (\ref{p1}) if the functions $f_j$ adopt
the following form
\begin{eqnarray}
f_j({\bm \sigma}_j)=\left\{
\begin{array}{cc} 1, & \sigma_{i\alpha}=\sigma_{i\beta}\ \ \forall \alpha,\beta\ni i\ \
\\
0, & \mbox{  otherwise.}\end{array}\right. \label{fg-ov}
\end{eqnarray}

\subsubsection{General Vertex model}
\label{subsubsec:Gen}

A general vertex model is determined by the weight function that can be represented in the
following form
\begin{eqnarray}
\label{VM-weight-gen}
 p_{\it gv}({\bm\sigma})= Z_{\it gv}^{-1}\prod\limits_{a\in X_{0}}f_a({\bm\sigma}_a),\ \  Z_{\it
 gv}=\sum\limits_{\bm \sigma} \prod\limits_{a\in
 X_{0}}f_a({\bm\sigma}_a),
\end{eqnarray}
where $a$ denotes a vertex in the model;  elementary spin is defined at the edge connecting two
neighboring vertexes,  $\sigma_{ab}$ for $b\in a$ and $a\in b$; ${\bm \sigma}_a$ stands for the
vector built from all $\sigma_{ab}$ where $b\in a$; ${\bm \sigma}$ is a particular configuration of
spins on all the edges. It is important to realize that with this notation we need to assume that
$\sigma_{ab}=\sigma_{ba}$.

A general vertex model turns into the orientable vertex model if the whole family of bits $\{ a\}$
is divided in two subfamilies that correspond to checks and bits, $a=i\oplus\alpha$, and
additionally for any bit/check the neighbors belong to the opposite families.

Therefore, for the example shown in Fig.~\ref{paths} the oriented vertex model and the the general
vertex model simply coincide as the graph allows partitioning in two parts. A simple example of a
general vertex model which does not correspond to any oriented case (the whole family of nodes is
not divisible in two groups) is given by an interconnected triad of vertexes with pair interaction:
$p\sim f_1(\sigma_{12},\sigma_{13})f_2(\sigma_{21},\sigma_{23})f_3(\sigma_{31},\sigma_{32})$.

\subsection{Loop series}
\label{subsec:Loop}

In this Subsection we state the main result of the paper for the
three models introduced above.

\subsubsection{General Vertex model}
\label{subsubsec:gv-loop}

We start with the general vertex model. The partition function of the general vertex model,
described by Eq.~(\ref{VM-weight-gen}) is exactly equal to
\begin{eqnarray}
Z_{\it gv}=Z_0\left(1+\sum\limits_{\it C} r_{\it gv}({\it C})\right),\quad
 r_{\it gv}({\it C})=\frac{\prod\limits_{a\in{\it C}}\mu_a}
{\prod\limits_{(ab)\in C}(1-m_{ab}^2)}, \label{gv_ser}
\end{eqnarray}
where summation goes over all allowed ${\it C}$ (marked) paths in
the graph associated with the model; $(ab)$ marks the edge on the
graph connecting nodes $a$ and $b$. The marked path is allowed to
branch at any node/vertex, however it cannot terminate at a node. We
refer to such a structure as a loop (it is actually some kind of a
generalized loop since branching is allowed; we use the shorter name
for convenience). $m_{ab}$ is the magnetization at the edge that
connects nodes $a$ and $b$. $\mu_a$ is the irreducible correlation
function at node $a$. The order of the correlation function is equal
to the number of marked nodes (nodes belonging to the marked path
${\it C}$) neighboring $a$. The bare partition function $Z_0$, the
magnetization $m_{ab}$, and the correlation functions $\mu_a$ are
calculated within the BP procedure, described by
Eqs.~(\ref{Gen_Pj},\ref{Gen-BP-conv},\ref{gen_prob_BP},\ref{gen-mu},\ref{gen-m}).

\subsubsection{Orientable Vertex model}
\label{subsubsec:ov-loop}

The general formula (\ref{gv_ser}) applied to the case of the nodes/vertexes partitioned into bits
and checks, reads
\begin{eqnarray}
 && Z_{\it ov}=Z_0\left(1+\sum_{\it C}r_{\it ov}({\it C})\right),\quad
 r_{\it ov}({\it C})=\frac{
 \left(\prod\limits_{i\in{\it C}}
 \mu_{i}\right)
 \left(\prod\limits_{\beta\in{\it C}}
 \mu_{\beta}\right)} {\prod\limits_{(i\alpha)\in C}(1-m_{i\alpha}^2)},
 \label{ov_ser}
\end{eqnarray}
where summation goes over all allowed (marked) paths ${\it C}$ in the graph associated with the
model. A marked path (generalized loop) is allowed to branch at any bit/check, however, it may not
terminate at a bit or check. In this case there are two types of irreducible correlation functions
associated with bits and checks, respectively, and one type of magnetization ( which is associated
with the edge that connects any bit with its neighboring check that necessarily belongs to the
loop) entering Eq.~(\ref{ov_ser}), all calcuated within BP and defined in
Eqs.~(\ref{ov-mu1},\ref{ov-mu2},\ref{ov-m}). Eq.~(\ref{ov_ser}) also follows directly from formulas
of Section \ref{subsec:Orientable}.

\subsubsection{Factor Graph model}
\label{subsubsec:fg-loop}

The decomposition of the partition function defined by  Eq.~(\ref{p1}) into a finite series has a
form:
\begin{eqnarray} &&
 Z_{\it fg}=Z_0\left(1+\sum\limits_{{\it C}}r_{\it fg}({\it C})
\right),\; r_{\it fg}({\it C})=\prod\limits_{i,\alpha\in {\it
C}}\mu_{\alpha}\mu_{i},
\label{Zseries}\\
 && \mu_{i}=\frac{(1-m_i)^{q_i-1}+(-1)^{q_i}(1+m_i)^{q_i-1}}{2(1-m_i^2)^{q_i-1}},\,
q_i=\sum\limits_{\alpha\in{\it C}}^{\alpha\ni i}1,
\label{m-fg}\\
&&
\mu_{\alpha}=\sum\limits_{\sigma_\alpha}b_\alpha(\sigma_\alpha)\prod\limits_{i\in{\it
C}}^{i\in\alpha}(\sigma_i-m_i),\; m_i=\sum\limits_{\sigma_i}
b_i(\sigma_i)\sigma_i. \label{mu-fg}
\end{eqnarray}
where summation goes over all allowed (marked) paths ${\it C}$
(generalized loops). They consist of sets of bits and checks so that
each of them has at least two distinct neighbors on the path. For
the aforementioned example there are fourteen allowed marked paths
(loops) shown in Fig.~(\ref{paths}) on the right. In
Eqs.~(\ref{Zseries}) $b_i(\sigma_i)$, $b_\alpha(\sigma_\alpha)$ and
$Z_0$ are beliefs (probabilities) defined on bits and checks and
partition function, respectively, calculated for the BP solution.
The BP solution for the model is described in detail in Section
\ref{sec:Fac}, see also Appendix \ref{app:Bethe}.

It is easy to verify that if Eq.~(\ref{fg-ov}) is assumed for the generalized vertex model,
Eq.(\ref{ov_ser}) turns exactly into Eq.~(\ref{Zseries}).  Indeed,  under condition of
(\ref{fg-ov}) the irreducible correlation functions at a check in the two formulae are exactly
equivalent. One derives
\begin{eqnarray} &&
\mu_{i}\to\int d\sigma_i p_i(\sigma_i)
(\sigma_i-m_i)^{q_i}=\frac{1-m_i^2}{2}\left[(1-m_i)^{q_i-1}+(-1)^{q_i}(1+m_i)^{q_i-1}\right],
\label{red1}\\ && \prod\limits_{\alpha\in i,{\it C}}^{i\in{\it
C}}(1-m_{i\alpha}^2)\to (1-m_i^2)^{q_i},\label{red2}
\end{eqnarray}
where the definition of the $d\sigma$ integration (summation) is given in Section
\ref{subsec:Orientable}, $p_i(\sigma_i)=(1+\sigma_i m_i)/2$ is the probability to find bit $i$ in
the state $\sigma_i$ within the BP solution, and $q_i$ is the connectivity degree of the bit $i$ at
the marked subgraph ${\it C}$, defined by Eq.~(\ref{m-fg}). All together the equivalence is
completely restored.

\section{Loop series derivation for the Vertex Models}
\label{sec:Ver}

\subsection{Vertex Model on Orientable Graphs}
\label{subsec:Orientable}

To introduce a representation that leads to the loop expansion it is convenient to introduce simple
integral calculus and discrete Fourier transform for functions $f(\sigma)$ of an Ising (spin)
variable. Note that ``integrals" here are nothing but sums over discrete sets, introduced solely to
simplify notations. A Fourier transform of $f(\sigma)$ is a function $\hat{{\cal F}}f(\pi)$, where
the corresponding momentum $\pi=\pm 1$ is also an Ising variable. The definitions and properties of
integrals and Fourier transform are as follows
\begin{eqnarray}
\label{integral} f(\sigma)&=&a+b\sigma; \;\;\; \int d\sigma
f(\sigma)=\sum_{\sigma=\pm
1}f(\sigma); \;\;\; \int d\sigma=2; \;\;\; \int \sigma d\sigma=0; \\
\label{fourier} \hat{{\cal F}}f(\pi)&=&\frac{1}{4}\int
d\sigma(1+\pi\sigma)f(\sigma); \;\;\;
\hat{{\cal F}}^{-1}g(\sigma)=\int d\pi(1+\pi\sigma)g(\pi); \nonumber \\
\hat{{\cal F}}(1)&=&\frac{1}{2}; \;\;\; \hat{{\cal
F}}(\sigma)=\frac{\pi}{2}; \;\;\; \hat{{\cal F}}^{-1}(1)=2; \;\;\;
\hat{{\cal F}}^{-1}(\pi)=2\sigma,
\end{eqnarray}
Denoting
\begin{eqnarray}
\label{define-F}
F_{\alpha}({\bm\sigma}_{\alpha})=f_{\alpha}({\bm\sigma}_{\alpha});
\;\;\; F_{j}({\bm\pi}_{j})=\hat{{\cal F}}f_{j}({\bm\pi}_{j}); \;\;\;
f_{j}({\bm\sigma})=\int
d{\bm\pi}_{j}F_{j}({\bm\pi}_{j})\prod_{\alpha\ni
j}(1+\pi_{j\alpha}\sigma_{\alpha j}),
\end{eqnarray}
where, $d{\bm \pi}_j=\prod_{\alpha\ni j}d\pi_{j\alpha}$, we can
represent the partition function in the form
\begin{eqnarray}
 \label{Z-intermediate} Z_{\it ov} =
 \int \prod\limits_\alpha d{\bm\sigma}_\alpha
 d{\bm\pi}_\alpha
 \left(\prod_{j}F_{j}({\bm\pi}_{j})\right)
 \left(\prod_{\alpha}F_{\alpha}({\bm\sigma}_{\alpha})\right)
 \left(\prod_{j\alpha}(1+\pi_{j\alpha}\sigma_{\alpha j})\right).
\end{eqnarray}
Here and below in this Subsection the index order in the definition of the discrete fields is
arbitrary, i.e. $\sigma_{i\alpha}=\sigma_{\alpha i}$. Our derivation of the loop expansion rests on
an important, yet very simple relation that can be easily verified directly:
\begin{eqnarray}
\label{edge-relation}
\frac{\cosh(\eta+\chi)(1+\pi\sigma)}{(\cosh\eta+\sigma\sinh\eta)(\cosh\chi+\pi\sinh\chi)}
=1+\left(\tanh(\eta+\chi)-\sigma\right)\left(\tanh(\eta+\chi)-\pi\right)\cosh^{2}(\eta+\chi).
\end{eqnarray}
Introducing two sets of parameters $\eta_{\alpha j}$ and $\chi_{j\alpha}$ that reside in the graph
edges we can make use of Eq.~(\ref{edge-relation}) to re-group the terms. This results in the
following expression for the partition function
\begin{eqnarray}
\label{Z-ready} && Z_{\it ov}=\bar{Z}_{\it gv}\int
d{\bm\sigma}d{\bm\pi}\prod_{j}P_{j}({\bm\pi}_{j})\prod_{\alpha}P_{\alpha}({\bm\sigma}_{\alpha})
\prod_{j\alpha}V_{j\alpha}\left(\sigma_{\alpha
j},\pi_{j\alpha}\right),\\ && \bar{Z}_{\it
gv}=\left(\prod_{j\alpha}\cosh\left(\eta_{\alpha
j}+\chi_{j\alpha}\right)\right)^{-1},
 \label{Zbar}\\ &&
 P_{j}({\bm\pi}_{j})=F_{j}({\bm\pi}_{j})\prod_{\alpha\ni j}
 \left(\cosh(\chi_{j\alpha})+\pi_{j\alpha}\sinh(\chi_{j\alpha})\right),
 \label{Pj}\\ &&
 P_{\alpha}({\bm\sigma}_{\alpha})=F_{\alpha}({\bm\sigma}_{j})\prod_{j\in\alpha}
\left(\cosh(\eta_{\alpha j})+\sigma_{\alpha j}\sinh(\eta_{\alpha j})\right), \label{Pa} \\
 && V_{j\alpha}\left(\sigma_{\alpha j},\pi_{j\alpha}\right)=1+\left(\tanh(\eta_{\alpha
j}+\chi_{j\alpha})-\sigma_{\alpha j}\right)\left(\tanh(\eta_{\alpha
j}+\chi_{j\alpha})-\pi_{j\alpha}\right)\cosh^{2}(\eta_{\alpha
j}+\chi_{j\alpha}).
\end{eqnarray}
The desired decomposition is obtained by expanding the $V$-terms followed by a local computation.
The parameters ${\bm\eta}$ and ${\bm\chi}$ are chosen using the criterion that skeletons
(subgraphs) with loose ends do not contribute to the decomposition. This can be achieved if the
parameters satisfy the following system of equations
\begin{eqnarray}
 \label{General-BP1}
 &&\int d{\bm\pi}_{j}\left(\tanh(\eta_{\alpha j}+\chi_{j\alpha})
 -\pi_{j\alpha}\right)P_{j}({\bm\pi}_{j})=0,\\
 && \label{General-BP2} \int d{\bm\sigma}_{\alpha}
 \left(\tanh(\eta_{\alpha j}+\chi_{j\alpha})-\sigma_{\alpha
 j}\right)P_{\alpha}({\bm\sigma}_{\alpha})=0.
\end{eqnarray}
The first equation in the system, Eq.~(\ref{General-BP1}), can actually be reduced by making use of
Eqs.~(\ref{integral}-\ref{define-F}) and (\ref{Pj}), to
\begin{eqnarray}
 && \label{General-BP1a}
 \int d{\bm\sigma}_{j}\left(\tanh(\eta_{\alpha j}+\chi_{j\alpha})
 -\sigma_{j\alpha}\right)\tilde{P}_{j}({\bm\sigma}_{j})=0,\\
 && \label{Ptilde}
 \tilde{P}_{j}({\bm\sigma}_{j})=f_{j}({\bm\sigma}_{j})\prod_{\alpha\ni j}
\left(\cosh(\chi_{j\alpha})+\sigma_{j\alpha}\sinh(\chi_{j\alpha})\right).
\end{eqnarray}
Combining Eqs.~(\ref{General-BP2},\ref{General-BP1a}) we derive
\begin{eqnarray} &&
 \frac{\exp\left[(\eta_{\alpha j}+\chi_{j\alpha})\sigma_{j\alpha}\right]}
 {\cosh\left[\eta_{\alpha j}+\chi_{j\alpha}\right]} =\sum\limits_{{\bm \sigma}_j\setminus\sigma_{j\alpha}}b_j^{(ov)}
 =\sum\limits_{{\bm \sigma}_\alpha\setminus\sigma_{\alpha j}}
 b_\alpha^{(ov)},
 \label{BPov}\\
 && b_j^{(ov)}=\frac{\tilde{P}_j({\bm\sigma}_j)}{\sum\limits_{{\bm \sigma}_j}
 \tilde{P}_j({\bm\sigma}_j)},\quad
    b_\alpha^{(ov)}=\frac{P_\alpha({\bm\sigma}_\alpha)}
 {\sum\limits_{{\bm \sigma}_\alpha}
 P_\alpha({\bm\sigma}_\alpha)},\label{p_ov_bp}
\end{eqnarray}
where it is assumed that $\sigma_{j\alpha}=\sigma_{\alpha j}$. Eq.~(\ref{BPov})constitutes the BP
system of equations, represented in terms of parameters $\eta$ and $\chi$. Eqs.~(\ref{p_ov_bp})
provide the BP expressions for the probabilities (beliefs) to observe the spin vector associated
with a bit/check in the corresponding states states.

A typical sum/integral, needed to calculate individual marked
path/diagram ${\it C}$ contribution, is reduced to the following
irreducible correlation functions  that should be computed within BP
\begin{eqnarray} &&
 \mu_{\alpha}=\int
 d{\bm\sigma}_{\alpha}b_\alpha^{\it (ov)}({\bm\sigma}_{\alpha})\prod\limits_{i\in \alpha,{\it C}}
\left(\sigma_{i \alpha}-m_{i \alpha}\right),
 \label{ov-mu1}\\
 &&
 \mu_{i}= \int
 d{\bm\sigma}_i b_i^{\it (gv)}({\bm\sigma}_i)\prod\limits_{\alpha\ni i; \alpha\in{\it C}}
 \left(\sigma_{i \alpha}-m_{i \alpha}\right),
 \label{ov-mu2}
\end{eqnarray}
where $m_{i\alpha}$ is the BP magnetization at the edge $i\alpha$
\begin{eqnarray}
m_{i\alpha}= \int d{\bm\sigma}_i b_i^{\it (ov)}({\bm\sigma}_{i})
\sigma_{i\alpha}
 = \int d{\bm\sigma}_\alpha b_\alpha^{\it (ov)}({\bm\sigma}_{\alpha})\sigma_{i\alpha}.
 \label{ov-m}
\end{eqnarray}

\subsection{General Vertex Model}
\label{subsec:GenVertex}

We are now in a position to consider the case of a general, not necessarily orientable, graph. The
loop expansion and the BP equations can be readily extended to this case. To derive the desired
loop decomposition we relax the condition $\sigma_{ab}=\sigma_{ba}$, i.e. we treat $\sigma_{ab}$
and $\sigma_{ba}$ as independent Ising variables. In complete analogy with the orientable case we
represent the partition function in a form:
\begin{eqnarray}
\label{Z-gen-graph} Z_{\it gv}=\int
d{\bm\sigma}\prod_{a}f_{a}({\bm\sigma}_{a})\prod_{bc}\frac{1+\sigma_{bc}\sigma_{cb}}{2}.
\end{eqnarray}
Note that for this representations the vectors ${\bm\sigma}_{a}$
become independent variables. Also in the product over $bc$ we
assume that each edge contributes only once. We further introduce a
parameter vector ${\bm\eta}$ with the components $\eta_{ab}$, all of
them being independent variables. Making use of
Eq.~(\ref{edge-relation}) we arrive at the following representation
for the partition function that is ready for the loop decomposition
\begin{eqnarray}
\label{Z-ready-gen} Z_{\it gv}&=&\bar{Z}_{\it gv}\int
d{\bm\sigma}\prod_{a}P_{a}({\bm\sigma}_{a})
\prod_{bc}V_{bc}\left(\sigma_{bc},\sigma_{cb}\right); \;\;\;
\bar{Z}_{\it
gv}=\left(\prod_{bc}2\cosh\left(\eta_{bc}+\eta_{cb}\right)\right)^{-1};
\nonumber
\\ P_{a}({\bm\sigma}_{a})&=&f_{a}({\bm\sigma}_{a})\prod_{b\in a}
\left(\cosh\eta_{ab}+\sigma_{ab}\sinh\eta_{ab}\right); \;\;\;
\label{Gen_Pj} \\
V_{bc}\left(\sigma_{bc},\sigma_{cb}\right)&=&1+\left(\tanh(\eta_{bc}+\eta_{cb})-\sigma_{bc}\right)
\left(\tanh(\eta_{bc}+\eta_{cb})-\sigma_{cb}\right)\cosh^{2}(\eta_{bc}+\eta_{cb}).
\end{eqnarray}
The BP equations for our general case have a form:
\begin{eqnarray}
\label{General-BP-gen} \int
d{\bm\sigma}_{a}\left(\tanh(\eta_{ab}+\eta_{ba})-\sigma_{ab}\right)P_{a}({\bm\sigma}_{a})=0.
\end{eqnarray}
To recast Eq.~(\ref{General-BP-gen}) in a stndard BP form we denote by ${\bm\eta}_{ab}$ the vector
with the components $\eta_{ac}$ with $c\in a$ and $c\ne b$, i.e.
${\bm\eta}_{a}=\left({\bm\eta}_{ab},\eta_{ab}\right)$. We also define a function
$\gamma({\bm\eta}_{ab})$ using the condition
\begin{eqnarray}
\label{gamma-cond} \int\prod_{c\in a}^{c\ne b}d\sigma_{ac}f_{a}({\bm\sigma}_{a})\prod_{c\in
a}^{c\ne
b}\left(\cosh\eta_{ac}+\sigma_{ac}\sinh\eta_{ac}\right)=\phi(\cosh\gamma+\sigma_{ab}\sinh\gamma).
\end{eqnarray}
The meaning of Eq.~(\ref{gamma-cond}) is as follows. The l.h.s. of the equation is a function of
the Ising variable $\sigma_{ab}$ and a function of ${\bm\eta}_{ab}$ (since by definition it does
not depend on $\eta_{ab}$). The r.h.s. constitutes a generic representation of such a function
provided $\phi$ and $\gamma$ are allowed to depend on ${\bm\eta}_{ab}$. Integrating (summing) over
$\sigma_{ab}$ in Eq.~(\ref{gamma-cond}) with and without the $\sigma_{ab}$ factor allows to
determine the function $\gamma({\bm\eta}_{ba})$ explicitly:
\begin{eqnarray}
\label{define-gamma} \tanh\gamma({\bm\eta}_{ab})=\frac{\int
d{\bm\sigma}_{a}\sigma_{ba}f_{a}({\bm\sigma}_{a})\prod_{c\in a}^{c\ne
b}\left(\cosh\eta_{ac}+\sigma_{ac}\sinh\eta_{ac}\right)}{\int
d{\bm\sigma}_{a}f_{a}({\bm\sigma}_{a})\prod_{c\in a}^{c\ne
b}\left(\cosh\eta_{ac}+\sigma_{ac}\sinh\eta_{ac}\right)}.
\end{eqnarray}
Multiplying Eq.~(\ref{gamma-cond}) with a factor $(\cosh\eta_{ab}+\sigma_{ab}\sinh\eta_{ab})$
yields
\begin{eqnarray}
\label{gamma-cond-2} \int\prod_{c\in a}^{b\ne b}d\sigma_{ca}P_{a}({\bm\sigma}_{a})
\phi\left(\cosh(\gamma+\eta_{ab})+\sigma_{ab}\sinh(\gamma+\eta_{ab})\right).
\end{eqnarray}
Comparing Eq.~(\ref{gamma-cond-2}) with Eq.~(\ref{General-BP-gen})
we arrive at $\sinh(\gamma-\eta_{ba})=0$. This allows to represent
the BP equations in a more conventional form
\begin{eqnarray}
\label{Gen-BP-conv} \eta_{ba}=\gamma\left({\bm\eta}_{ab}\right).
\end{eqnarray}
Calculated within BP, the probability of finding the whole family of
edges connected to node $a$ in the state ${\bm \sigma}_a$ is
\begin{eqnarray}
 b_a^{\it (gv)}({\bm \sigma}_a)=\frac{P_{a}({\bm\sigma}_{a})}{\int d{\bm
 \sigma}_a
 P_a({\bm\sigma}_a)}.
\label{gen_prob_BP}
\end{eqnarray}
In the general vertex model case a typical integral (sum), needed to
take to calculate a diagram contribution for a generalized loop
${\it C}$, is reduced to the corresponding irreducible correlation
functions of the spin variables computed within BP:
\begin{eqnarray}
 \mu_{a}= \int
 d{\bm\sigma}_{a}b_a^{\it (gv)}({\bm\sigma}_{a})\prod\limits_{b\in a,{\it C}}
 \left(\sigma_{ab}-m_{a b}\right),
 \label{gen-mu}
\end{eqnarray}
where $m_{ab}$ is the magnetization at the edge $(ab)$ calculated
within BP
\begin{eqnarray}
 m_{ab}=
 \int d{\bm\sigma}_{a} b_a^{\it (gv)}({\bm\sigma}_{a}) \sigma_{ab}.
 \label{gen-m}
\end{eqnarray}
The final and most general expression Eq.~(\ref{gv_ser}) emerges in
the result of direct calculation of the (generalized) loop
contributions making use of
Eqs.~(\ref{Z-ready-gen},\ref{Gen-BP-conv},\ref{gen-mu},\ref{gen-m}).

\section{Loop series derivation for the factor-graph model}
\label{sec:Fac}

\subsection{Integral representation for the Factor-graph  model}
\label{subsec:Int}

We aim to derive a convenient integral representation for the
statistical model (\ref{p1}). As a first step we introduce two
statistically independent sets of discrete random variables: the
original $\{\sigma_i\}$, and the additional factor-variable
counterpart $\{\pi_\alpha\}$, where each $\pi_\alpha$ is a vector
consisting of $q_\alpha$ scalar components,  each a discrete random
variable, and $q_\alpha$ is the degree of connectivity of the
corresponding factor node. For the example represented by
Fig.~(\ref{paths}) we have $\pi_a=(\pi_a^{(1)},\pi_a^{(2)}),
\pi_b=(\pi_b^{(1)},\pi_b^{(2)}),
\pi_c=(\pi_c^{(1)},\pi_c^{(3)},\pi_c^{(4)}),
\pi_d=(\pi_c^{(3)},\pi_c^{(4)})$ where $\pi_{a,b,c,d}^{(i)}=\pm 1$.
Using such a representation the partition function of Eq.~(\ref{p1})
can be rewritten as
\begin{eqnarray}
 Z\sim\sum\limits_{\{\pi_\alpha\}}
 \left[\prod\limits_\alpha
 f_\alpha(\pi_\alpha)\right]\prod\limits_i
 \left[\sum\limits_{\sigma_i}\prod\limits_{\alpha\ni
 i}\delta\left(\sigma_i,\pi_\alpha^{(i)}\right)\right]
 ,\label{Z1}
\end{eqnarray}
where the product over $i$ is taken over the bits connected to more then one factor nodes. Under
condition that all discrete scalars $\pi^{(i)}_\alpha$ belong to the binary alphabet the expression
on the rhs of Eq.~(\ref{Z1}) can be rewritten as
\begin{eqnarray} &&
 \sum\limits_{\sigma_i}\prod\limits_{\alpha\ni i}\delta\left(\sigma_i,\pi^{(i)}_\alpha\right)
 \sim\int_{{\bm C}_{i}}d{\bm\chi}_{i}
\exp\left(\sum\limits_{\alpha\ni
i}\chi_{i\alpha}\pi_\alpha^{(i)}\right)
 \left[\sum\limits_{\sigma_i}\exp\left(\frac{1}{q_i-1}\ \sigma_i\sum\limits_{\alpha\ni
 i}\chi_{i\alpha}\right)\right]^{1-q_i}, \label{Z2}
\end{eqnarray}
where $q_i>1$ is the degree of connectivity of the bit $i$, and ${\bm\chi}_{i}$ is a vector with
the components $\chi_{i\alpha}$, where $\alpha\ni i$. Integration goes over a $q_{i}$-dimensional
cycle ${\bm C}_{i}=\prod_{\alpha\ni i}C_{i\alpha}$ that constitutes a cartesian product of $q_{i}$
contours in the complex plane: $C_{j\alpha}$ connects the points $z_{j\alpha}$ and
$z_{j\alpha}+2\pi i\left(q_{j}-1\right)$ in an arbitrary way however such that the contour does not
go through the point of formal singularity of the integrand in Eq.~(\ref{Z2}). It is
straightforward to check that the result does not depend on the particular choice of the reference
points $z_{i\alpha}$. The multidimensional integration contour can be defined in way (in the sense
of passing the multidimensional pole manifold) that the integral representation is exact, yet its
deformation that reaches the saddle point does not involve the pole manifold. This is confirmed
indirectly by identical exact loop expansions that originate from the integral representation and
it discrete counterpart. Note also that the integral representation is obviously not unique and the
specific choice of the representation is dictated by our desire to find one that guarantees
emergence of the BP in the saddle-point approximation applied to the integral. Below in Section
\ref{subsec:SP} we will verify, indeed, that Eq.~(\ref{Z2}) obeys the desired saddle-point
property.

Substituting Eq.(\ref{Z2}) into Eq.~(\ref{Z1}) one derives
\begin{eqnarray} &&
 Z\sim \int \left[\prod\limits_i\prod\limits_\alpha d\chi_{i\alpha}\right]
 \prod\limits_i\left[\sum\limits_{\sigma_i}
 \exp\left(\frac{1}{q_i-1}\ \sigma_i\sum\limits_{\alpha\ni i}
 \chi_{i\alpha}\right)\right]^{1-q_i}
 \prod\limits_\alpha\left[\sum\limits_{\pi_\alpha}\left(
 f_\alpha(\pi_\alpha)\exp\left(\sum\limits_{i\in\alpha}
 \pi_\alpha^{(i)}\chi_{i\alpha}\right)\right)\right]
 \label{Z3}\\ &&
 =\int \left[\prod\limits_i\prod\limits_\alpha d\chi_{i\alpha}\right]
 \left(\prod\limits_i\exp\left[-Q_i({\bm \chi})\right]\right)
 \left(\prod\limits_\alpha\left[\sum\limits_{\pi_\alpha}
 \exp\left[-Q_\alpha({\bm\chi})\right]\right]\right)= \int \left[\prod\limits_i\prod\limits_\alpha
 d\chi_{i\alpha}\right] \exp\left[-{\cal
 S}_0(\chi)\right].
 \label{Z30}
\end{eqnarray}

\subsection{Relation to the Bethe variational approach}
\label{subsec:Bethe}

The expression of Eq.~(\ref{Z3}) is compact and already constitutes a good starting point for
further, e.g. saddle point, analysis. Meantime, for the purpose of establishing a relation to the
Bethe free energy approach of \cite{05YFW} and for some further applications we introduce the
following auxiliary integrations
\begin{eqnarray}
 && 1\sim\int\left[\prod\limits_i
 \prod\limits_{\sigma_i}d\varphi_i (\sigma_i)d\bar{\varphi}_i(\sigma_i)\right]
 \exp\left[\sum\limits_i
 \sum\limits_{\sigma_i}
 \bar{\varphi}_i(\sigma_i)
 \left(\varphi_i(\sigma_i)-\sigma_i\sum\limits_{\alpha\ni i}\chi_{i\alpha}\right)\right],
 \label{1i}\\
 && 1\sim
 \int\left[\prod\limits_\alpha
 \prod\limits_{\pi_\alpha}d\psi_\alpha(\pi_\alpha)d\bar{\psi}_\alpha(\pi_\alpha)\right]
 \exp\left[\sum\limits_\alpha
 \sum\limits_{\pi_\alpha}\bar{\psi}_\alpha(\pi_\alpha)
 \left(\psi_\alpha(\pi_\alpha)+\ln f_\alpha(\pi_\alpha)+\sum\limits_{i\in\alpha}
 \pi_\alpha^{(i)}\chi_{i\alpha}\right)\right],\label{1a}
\end{eqnarray}
in the rhs of Eq.~(\ref{Z3}). After some obvious manipulations we arrive at
\begin{eqnarray} &&
 Z\sim \int \left[\prod\limits_i\prod\limits_\alpha d\chi_{i\alpha}\right]
 \left[\prod\limits_i
 \prod\limits_{\sigma_i}d\varphi_i(\sigma_i)d\bar{\varphi}_i(\sigma_i)\right]
 \left[\prod\limits_\alpha
 \prod\limits_{\pi_\alpha}d\psi_\alpha(\pi_\alpha)d\bar{\psi}_\alpha(\pi_\alpha)\right]
 \exp\left[-{\cal S}\right], \label{Z3a}\\
 && {\cal S}= \sum\limits_i\left[-\sum\limits_{\sigma_i}\bar{\varphi}_i(\sigma_i)
 \left(\varphi_i(\sigma_i)-\sigma_i\sum\limits_{\alpha\ni
 i}\chi_{i\alpha}\right)+
 (q_i-1)\ln\left(\sum\limits_{\sigma_i}\exp\left(\frac{\varphi_i(\sigma_i)}{q_i-1}\right)\right)
 \right]
 \nonumber\\ &&
 -\sum\limits_\alpha\left[
 \sum\limits_{\pi_\alpha}\bar{\psi}_\alpha(\pi_\alpha)
 \left(\psi_\alpha(\pi_\alpha)+\ln f_\alpha(\pi_\alpha)+\sum\limits_{i\in\alpha}
 \pi_\alpha^{(i)}\chi_{i\alpha}\right)
 +\ln\left(\sum\limits_{\pi_\alpha}\exp\left(-\psi_\alpha(\pi_\alpha)\right)\right)
 \right]. \label{Z3b}
\end{eqnarray}

Evaluating the integral over $\varphi_i(\sigma_i)$, $\psi_\alpha(\pi_\alpha)$ within the
saddle-point approximation we obtain
\begin{eqnarray}
 && \bar{\varphi}_i(\sigma_i)= \frac{\exp\left(
 \varphi_i^{\small{\mbox{(sp)}}}(\sigma_i)/(q_i-1)\right)}{\sum\limits_{\sigma_i}\exp
 \left(\varphi_i^{\small{\mbox{(sp)}}}(\sigma_i)/(q_i-1)\right)}= (z^{\small{\mbox{(sp)}}}_i)^{-1}
 \exp\left( \varphi^{\small{\mbox{(sp)}}}_i(\sigma_i)/(q_i-1)\right)
 , \label{SP1}\\ &&
 \bar{\psi}_\alpha(\pi_\alpha)= \frac{\exp\left(-\psi_\alpha^{\small{\mbox{(sp)}}}(\pi_\alpha)\right)}
 {\sum\limits_{\pi_\alpha}\exp\left(-\psi_\alpha^{\small{\mbox{(sp)}}}(\pi_\alpha)\right)}
 =(z_\alpha^{\small{\mbox{(sp)}}})^{-1}\exp\left(-\psi^{\small{\mbox{(sp)}}}_\alpha(\pi_\alpha)\right).
 \label{SP2}
 \end{eqnarray}
Expressing $\varphi_i(\sigma_i)$, $\psi_\alpha(\pi_\alpha)$ in terms of
$\bar{\varphi}_i(\sigma_i)$, $\bar{\psi}_\alpha(\pi_\alpha)$ according to Eq.~(\ref{SP1},\ref{SP2})
and substituting the result into the effective action (\ref{Z3b}) we find
\begin{eqnarray} &&
 {\cal S}^{\small{\mbox{(sp)}}}= -\sum\limits_\alpha\sum\limits_{\pi_\alpha} \bar{\psi}_\alpha
 (\pi_\alpha)\ln f_\alpha(\pi_\alpha)
 \nonumber\\ &&
 +\sum\limits_\alpha\sum\limits_{\pi_\alpha}\bar{\psi}_\alpha
 (\pi_\alpha)\ln \bar{\psi}_\alpha(\pi_\alpha)
 -\sum\limits_i\sum\limits_{\sigma_i}(q_i-1)\bar{\varphi}_i
 (\sigma_i)\ln \bar{\varphi}_i(\sigma_i)
 \nonumber\\ &&
 +\sum\limits_i\sum\limits_{\alpha\ni i}\chi_{i\alpha}\sum\limits_{\sigma_i}\sigma_i
 \left(\bar{\varphi}_i(\sigma_i)-\sum\limits_{\pi_\alpha\setminus\sigma_i}
 \bar{\psi}_\alpha(\pi_\alpha)\right)
 -\sum\limits_\alpha \ln z_\alpha^{\small{\mbox{(sp)}}}+
 \sum\limits_i(q_i-1)\ln z_i^{\small{\mbox{(sp)}}}.\label{Seff}
\end{eqnarray}
The saddle-point (in $\psi$ and $\varphi$) solution (\ref{SP1},\ref{SP2}) is highly degenerate:
there is a freedom in imposing a constraint per any bit $i$ and per any factor-node $\alpha$.
Moreover, the integrand in Eq.~(\ref{Z3a},\ref{Z3b}) is invariant under the transformations:
\begin{eqnarray}
\psi_\alpha(\pi_\alpha)\to \psi_\alpha(\pi_\alpha) +c_\alpha,\quad
\varphi_i(\sigma_i)\to\varphi_i(\sigma_i)+c_i. \label{cc}
\end{eqnarray}
Fixing the values of $\sum_{\sigma_i} \varphi_i(\sigma_i)$ and
$\sum_{\pi_\alpha}\psi_\alpha(\pi_\alpha)$, introducing the shifts (\ref{cc}) into
Eq.~(\ref{Z3a},\ref{Z3b}) and integrating with respect to $c_i$,$c_\alpha$ one arrives at the
normalization constraints
\begin{eqnarray}
 \sum\limits_{\sigma_i}\bar{\varphi}_i(\sigma_i)=1,\quad
 \sum\limits_{\pi_\alpha}\bar{\varphi}_i(\pi_\alpha)=1, \label{norm}
\end{eqnarray}
that are dynamically imposed, i.e. they are present in the integrand of Eq.(\ref{Z3a}) as products
of the corresponding sets of $\delta$-functions. A convenient choice of $\sum_{\sigma_i}
\varphi_i(\sigma_i)$ and $\sum_{\pi_\alpha}\psi_\alpha(\pi_\alpha)$ constraints is the one that
makes $z_i^{\small{\mbox{(sp)}}}=z_\alpha^{\small{\mbox{(sp)}}}=1$. As a result the last two terms
on the rhs of Eq.~(\ref{Seff}) disappear and the equivalence between the effective action
(\ref{Seff}) and the Bethe free energy of \cite{05YFW} (see also Appendix \ref{app:Bethe}) becomes
clear.

\subsection{Belief-propagation as a saddle-point}
\label{subsec:SP}

Looking for the saddle-point configurations of the auxiliary fields $\bar{\varphi}_i(\sigma_i)$,
$\bar{\psi}_\alpha(\pi_\alpha)$, $\varphi_i(\sigma_i)$, $\psi_\alpha(\pi_\alpha)$ and
$\chi_{i\alpha}$ that dominate the contribution to the integral in Eq.~(\ref{Z3a}), and thus
setting the corresponding partial derivatives of ${\cal S}$ to zero, we obtain in addition to
Eqs.~(\ref{SP1},\ref{SP2}) the following saddle-point equations
\begin{eqnarray}
 &&
 \varphi_i^{\small{\mbox{(sp)}}}(\sigma_i)=\sigma_i\sum\limits_{\alpha\ni
 i}\chi_{i\alpha}^{\small{\mbox{(sp)}}},\label{SP3}\\ &&
 \psi_\alpha^{\small{\mbox{(sp)}}}(\pi_\alpha)+\ln f_\alpha(\pi_\alpha)+\sum\limits_{i\in\alpha}
 \pi_\alpha^{(i)}\chi_{i\alpha}^{\small{\mbox{(sp)}}}=0,\label{SP4}\\ &&
 \sum\limits_{\sigma_i}\sigma_i\bar{\varphi}_i^{\small{\mbox{(sp)}}}(\sigma_i)=\sum\limits_{\pi_\alpha}
 \pi_\alpha^{(i)}\bar{\psi}_\alpha^{\small{\mbox{(sp)}}}(\pi_\alpha). \label{SP5}
\end{eqnarray}
This system of Eqs.~(\ref{SP1},\ref{SP2},\ref{SP3}-\ref{SP5}) is identical to the BP system of
equation derived via variation of the Bethe free energy \cite{05YFW} (see also Appendix
\ref{app:Bethe}). The relation between the corresponding fields is as follows:
$\bar{\varphi}^{\small{\mbox{(sp)}}}_i(\sigma_i)\leftrightarrow b_i(\sigma_i)$,
$\bar{\psi}^{\small{\mbox{(sp)}}}_\alpha(\pi_\alpha)\leftrightarrow b_\alpha(\pi_\alpha)$,
$\chi_{i\alpha}^{\small{\mbox{(sp)}}}\sigma_i\leftrightarrow \lambda_{i\alpha}(\sigma_i)$,
$\chi_{i\alpha}^{\small{\mbox{(sp)}}}\leftrightarrow \sum_{\beta\ni
i}^{\beta\neq\alpha}\eta_{i\beta}$. Normalization constraints (\ref{norm}) are obviously satisfied
in Eqs.~(\ref{SP1},\ref{SP2}). The consistency constraint (\ref{cons}) is equivalent to
(\ref{SP5}).

Eqs.~(\ref{SP1},\ref{SP2},\ref{SP3}-\ref{SP5}) results in
\begin{eqnarray}
 \frac{\sum_{\pi_\alpha\setminus\sigma_i}f_\alpha(\pi_\alpha)
 \exp\left(\sum_{j\in\alpha}\pi_\alpha^{(j)}\chi_{j\alpha}^{\small{\mbox{(sp)}}}\right)}{
 \sum_{\pi_\alpha}f_\alpha(\pi_\alpha)\exp\left(\sum_{j\in\alpha}
 \pi_\alpha^{(j)}\chi_{j\alpha}^{\small{\mbox{(sp)}}}\right)} =\frac{\exp\left(\sigma_i\sum_{\beta\ni
 i}\chi_{i\beta}^{\small{\mbox{(sp)}}}/(q_i-1)\right)}{\sum_{\sigma_i}\exp\left(\sigma_i\sum_{\beta\ni
 i}\chi_{i\beta}^{\small{\mbox{(sp)}}}/(q_i-1)\right)}.
 \label{chi_eq}
\end{eqnarray}
that can also be derived directly from Eq.~(\ref{Z3}). Left-hand-side or right-hand-side of
Eq.~(\ref{chi_eq}) gives saddle-point, BP expression for $\bar{\varphi}_i(\sigma_i)$ -- the
probability to observe spin at the bit $i$ in the state $\sigma_i$.

The set of Eqs.~(\ref{SP1},\ref{SP2},\ref{SP3}-\ref{Z3}) coincides with the one derived as an
extremum condition for the Bethe-Free energy \cite{05YFW} (see also Appendix \ref{app:Bethe}).
Iterative solution of these nonlinear equations reproduces the famous Belief Propagation algorithm
for efficient yet sub-optimal solution of the inference problem.

The saddle-point approximation for the partition function
\begin{eqnarray}
 Z_0\sim \exp\left[-{\cal S}_0\left(\chi^{(sp)}\right)\right],
 \label{Z0}
\end{eqnarray}
is expressed in terms of the effective action ${\cal S}_0$ defined in (\ref{Z3},\ref{Z30}). Note,
that there may be more than one realizable (correspondent to real valued $\chi^{(sp)}$) solution of
the saddle-point (BP) system of equations.

For the purpose of further applications let us also introduce the magnetization and irreducable
correlation function (defined at two bits neighboring the same check), both defined within the
saddle-point BP approximation
\begin{eqnarray}
 && m_i=\sum\limits_{\sigma_i}\sigma_i\bar{\varphi}_i^{(sp)}(\sigma_i),\label{mag1}\\
 && \mbox{for}\ \ i,j\in\alpha:\ \ \mu_{ij}=
 \sum\limits_{\sigma_\alpha}(\sigma_i-m_i)(\sigma_j-m_j)\bar{\psi}_\alpha^{(sp)}(\sigma_\alpha)
 .\label{mu1}
\end{eqnarray}

\subsection{Gaussian correction to the saddle-point approximation}
\label{subsec:Gauss}

To calculate the correction to the zero-order saddle-point approximation we need to expand the
effective action ${\cal S}_0$ in $\chi_{i\alpha}$ around $\chi^{(\mbox{\small sp})}$ to the second
order. According to the definition of the saddle-point the first-order term in the expansion is
exactly zero. If the second-order expansion is sufficient, i.e. the higher-order terms are much
smaller with respect to some parameter (the exact origin of the expansion parameter will be
verified and discussed later), we shift the multi-dimensional integration contour that enters
Eq.~(\ref{Z30}) in the space of complex $\chi$-fields to go exactly through the saddle-point. The
next task is to calculate corrections that originate from the vicinity of the saddle-point. At the
saddle-point we choose the local orientation of the integration contour in the steepest descent
way. Note, that the steepest descent at a saddle-point may go along imaginary or real direction.
Finally the integral in Eq.~(\ref{Z30}) is approximated by a Gaussian integral. The result of the
Gaussian integration in Eq.~(\ref{Z30}), which leads to a correction the saddle-point term,
becomes
\begin{eqnarray}
\sim\exp\left[-{\cal S}_0-{\cal S}_1\right],\quad {\cal
S}_1=\frac{1}{2}\ln\left|\det\left[\hat{\Lambda}\right]\right|,\quad
\Lambda_{i\alpha;j\beta}(\chi^{(\mbox{\small sp})})\equiv\left.
\frac{\partial^2{\cal
 S}_0(\chi)}{\partial \chi_{i\alpha}\partial \chi_{j\beta}}
 \right|_{\chi=\chi^{(\mbox{\small sp})}},\label{Gauss_corr}
\end{eqnarray}
where all the expressions are taken at $\chi^{(\mbox{\small sp})}$.

Calculating the matrix of the second-order derivatives directly from Eq.~(\ref{Z3}) one arrives at
\begin{eqnarray}
 && \forall \quad i\in\alpha:\quad \Lambda_{i\alpha;i\alpha}(\chi^{(\mbox{\small sp})})=
 \frac{2-q_i}{q_i-1}\left[1-m_i^2\right],
  \label{F01}\\ && \forall\quad \alpha,\beta\ni i,\quad \alpha\neq\beta:\quad
\Lambda_{i\alpha;i\beta}(\chi^{(\mbox{\small sp})})=\frac{1-m_i^2}{q_i-1},\label{F02}\\
 && \forall\quad i,j\in\alpha,\quad i\neq j:\quad
 \Lambda_{i\alpha;j\alpha}(\chi^{(\mbox{\small sp})})=-\mu_{ij},\label{F03}
\end{eqnarray}
where all matrix elements between the pairs/links $\{i,\alpha\}$ and $\{j,\beta\}$ sharing neither
a common bit nor a common factor/check node are zero. In Eq.~(\ref{F01},\ref{F02},\ref{F03}) $m_i$
and $\mu_{ij}$  are the magnetizations and irreducible correlation functions, respectively, that
are calculated within the saddle-point (BP) approximation [see Eqs.~(\ref{mag1},\ref{mu1})]. Direct
calculation of the Gaussian integrals based on the problem-specific information on the BP
saddle-point solutions and the quadratic form matrix (\ref{F01},\ref{F03}), the latter also
depending on the bare saddle-point solutions, provides a straightforward algebraic way of computing
the partition function, magnetization, and any other spin-relate objects within the Gaussian
approximation.

A potential difficulty in the evaluation of Gaussian integrals
originates from a relatively complex structure of the  matrix
$\hat{\Lambda}$. Specifically, the off-diagonal (\ref{F03}) term
proportional to the irreducible pair correlation function induces
coupling between different blocks related to the corresponding bits.
In the following Subsection we analyze the Gaussian integrals
perturbatively by expanding the off-diagonal term in an infinite
series.

We will show that the only surviving terms in this expansion, after the Gaussian integrations are
performed, correspond to loops inn the factor graphs. We will demonstrate that the actual expansion
parameter is the product over the loop of the terms $\mu_{ij}/(1-m_i^2)$, evaluated at the saddle
point.

\subsubsection{Loop expansion in the Gaussian case}
\label{subsubsec:loopGauss}

We start with introducing a convenient notation for the Gaussian integration
\begin{eqnarray}
 \langle A\rangle_{\mbox{\small bd}}\equiv
 \frac{\int \left(\prod\limits_{p,\gamma}^{\gamma\in p}
 d\zeta_{p\gamma}\right) A
 \exp\left[-\frac{1}{2}\zeta_{n\beta}\Lambda^{(\mbox{\small bd})}_{n\beta;p\gamma}
 \zeta_{p\gamma}\right]}{\int \left(\prod\limits_{p,\gamma}^{\gamma\in p}
 d\zeta_{p\gamma}\right)
 \exp\left[-\frac{1}{2}\zeta_{n\beta}\Lambda^{(\mbox{\small bd})}_{n\beta;p\gamma}
 \zeta_{p\gamma}\right]},
 \label{G2}
\end{eqnarray}
where $\hat{\Lambda}^{(\mbox{\small bd})}$ is the block (bit)-diagonal part
$\hat{\Lambda}^{(\mbox{off})}$ of $\hat{\Lambda}$ defined by Eqs.~(\ref{F01},\ref{F03}) with the
off-block-diagonal part of $\hat{\Lambda}$, described by Eq.~(\ref{F02}) being ignored.

\begin{figure} 
\includegraphics[width=0.3\textwidth]{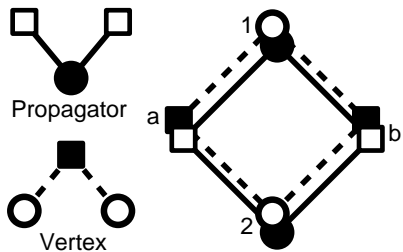}
\caption{Gaussian approximation about the BP saddle-point. Diagrams
for the propagator and the vertex are shown on the left. The right
plot illustrates a loop contribution. The example corresponds to a
loop in the model shown in the upper left corner of
Fig.~(\ref{paths}). A leg of a vertex (dashed line) should pair with
a leg of a propagator (solid line). No unpaired legs are allowed.}
\label{loop}
\end{figure}

An important object $P_{i\alpha;j\beta}=\langle \zeta_{i\alpha}\zeta_{j\beta}\rangle_{\mbox{\small
bd}}$ in the expansion with respect to $\hat{\Lambda}^{(\mbox{off})}$ is will be refered to as a
propagator following the traditional physics jargon of the Feynmann diagram expansion. It follows
from Eq.~(\ref{F01},\ref{F02}) that the only non-zero component of the propagator is represented by
\begin{eqnarray}
 \alpha\neq\beta,\quad \alpha,\beta\ni i:\quad
 P_{i\alpha;i\beta}=\frac{1}{1-m_i^2}.
 \label{P1}
\end{eqnarray}
Note that the propagator has an interesting ``fermionic-repulsive" feature: for a fixed bit $i$ it
is strictly zero for coinciding factor/check indices, i.e. $P_{i\alpha;i\alpha}=0$. In addition to
the ``propagator", the off-block-diagonal term is represented by a ``vertex". The vertex term is
nonzero only for
\begin{eqnarray}
 i\neq j,\quad i,j\in \alpha:\quad
 V_{i\alpha;i\beta}=\mu_{ij}.
 \label{P1}
\end{eqnarray}
It is also convenient to introduce a graphical notations for both the propagator and the vertex,
(see the left part of Fig.~\ref{loop}).

The correction to the partition function adopts the following form
\begin{eqnarray}
 Z_g=\Biggl\langle
 \exp\left[-\frac{1}{2}\zeta_{n\beta}\Lambda^{(\mbox{\small
 off})}_{n\beta;p\gamma}
 \zeta_{p\gamma}\right]\Biggr\rangle_{\mbox{\small bd}}= \sum\limits_{n=0}^\infty \frac{1}{2^n n!}\Biggl\langle
 \left[ -\zeta_{l\beta}\Lambda^{(\mbox{\small
 off})}_{l\beta;p\gamma}\zeta_{p\gamma}\right]^n\Biggr\rangle_{\mbox{\small
 bd}},
 \label{Zq}
\end{eqnarray}
where the full partition function of the model is approximated as
$Z\approx Z_0 Z_g$. Each term in the sum on the rhs of
Eq.~(\ref{Zq}) can be represented by a diagram. For an $n$-th order
term the diagram contains $n$-vertices. The Gaussian integration
that corresponds to each term is performed in the following way. We
first consider all possible Wick decompositions of the product of
$2n$ $\zeta$-terms in $n$ pairs. Each pair in the product results in
the corresponding propagator. The $n$-th order term on the rhs of
Eq.~(\ref{Zq}) naturally decomposes into a sum of $n(n-1)$ terms
each equal to a product of $n$-propagators and $n$-vertexes. The key
observation is that only very few of the terms survive due to the
specific structure of the propagators (\ref{P1}) and the vertexes
(\ref{V1}). Indeed, only those terms do not vanish that consist of
the propagators and vertexes, coupled through their legs and forming
a loop in the model factor graph. The structure is illustrated in
Fig.~\ref{loop}. It is common for the Feynman diagrammatic
techniques that the natural object is the logarithm $\ln(Z_g)$ of
the partition function, since only connected diagrams, i.e. the ones
that cannot be decomposed into a product of other diagrams,
contribute to the object. This results in
\begin{eqnarray}
 \ln(Z_g)=\sum\limits_{\it C}
 \frac{\prod\limits_{\alpha\in {\it C}}^{i,j\in\alpha}\mu_{ij}}{\prod\limits_{i\in {\it C}}(1-m_i^2)}, \label{Zq1}
\end{eqnarray}
where loops ${\it C}$ are defined as closed directed self-avoiding
paths in the model factor graph that pass from a bit to a
factor/check and then from a factor/check to a bit, etc, in such a
way that returns from a check that belongs to the path back to the
preceding bit are not allowed. An example of a loop is shown in
Fig.~\ref{loop}.

\subsection{Loop calculus via Integral Representation}
\label{subsec:LoopInt}

The Gaussian fluctuations analysis is justified only if the
higher-order (third-, fourth- , etc) corrections to the Gaussian
approximation are small compared to the major saddle-point and
Gaussian contributions, and the expansion is controlled by some
parameter. Jumping ahead we know that the loop expansion exposed by
the Gaussian approximation is the correct one, in the sense that the
connected loops contributions (no branching) provides the leading
correction with respect to the branching parameter. However, to see
how this general loop expansion actually works we need to expand the
effective action to all orders around the saddle point and classify
an infinite number of perturbative terms, which seems a nightmare.

Fortunately,  there is a way out of this technical problem that allows to account for all-order
corrections simultaneously. The method is based on introducing a set of new variables
$\zeta_{i\alpha}\equiv \chi_{i\alpha}-\chi_{i\alpha}^{(sp)}$ followed by explicit decomposition of
the integrand in Eq.~(\ref{Z3}) as a product of two non-Gaussian (with respect to the fields
$\zeta_{i\alpha}$) terms, which are diagonal in the bit and factor/check representations,
respectively:
\begin{eqnarray} &&
 Z\sim \int \left[\prod\limits_i\prod\limits_\alpha d\zeta_{i\alpha}\right]
 \prod\limits_{i} P_i\left(\zeta;\chi^{(sp)}\right)
 \prod\limits_\alpha
 V_\alpha\left(\zeta;\chi^{(sp)}\right),
 \label{I1}\\ &&
 P_i\equiv\frac{\prod\limits_{\alpha\ni i}\left[\sum\limits_{\pi_\alpha} f_\alpha(\pi_\alpha)
 \exp\left(\sum\limits_{j\in\alpha}\pi_\alpha^{(j)}\chi_{j\alpha}^{(sp)}\right)
 \exp\left(\pi_\alpha^{(i)}\zeta_{i\alpha}\right)\right]}
 {\left[\cosh\left(\sum\limits_{\alpha\ni
 i}(\chi_{i\alpha}^{(sp)}+\zeta_{i\alpha})/(q_i-1)\right)\right]^{q_i-1}},
 \label{P1}\\ &&
 V_\alpha\equiv \frac{
 \sum\limits_{\pi_\alpha}f_\alpha(\pi_\alpha)\exp\left(\sum\limits_{j\in\alpha}\pi^{(j)}_\alpha
 \left[\chi_{i\alpha}^{(sp)}+\zeta_{i\alpha}\right]\right)}
 {\prod\limits_{i\in\alpha}\left[\sum\limits_{\pi_\alpha}
 f_\alpha(\pi_\alpha)
 \exp\left(\sum\limits_{j\in\alpha}\pi_\alpha^{(j)}\chi_{j\alpha}^{(sp)}\right)
 \exp\left(\pi_\alpha^{(i)}\zeta_{i\alpha}\right)\right]}.
 \label{V1}
\end{eqnarray}
Here we introduced the same factor to the numerator of $P$ and
denominator of $V$, respectively, that is local both in bit and
check representations. The rational behind such a decomposition
choice is to ensure that in the of case of the Gaussian approximate
perturbative analysis of the effective action all correlations
within a block associated with the same bit are all included in the
``propagator" term $P$, while the inter-bit interaction appear only
in the ``vertex'' counterpart $V$.

We further introduce a set of convenient notations that will allow
to substantially simplify calculations:
\begin{eqnarray} &&
 \forall\ \alpha\neq\beta:\ \langle
 A(\pi_\alpha)B(\pi_\beta)\rangle_\pi=\langle
 A(\pi_\alpha)\rangle_{\pi_\alpha}\langle
 B(\pi_\beta)\rangle_{\pi_\beta},\label{pi_av1}\\ &&
 \Biggl\langle A(\pi_\alpha)\Biggr\rangle_{\pi_\alpha}\!\!\!\!\!\equiv\!
 \frac{\sum\limits_{\pi_\alpha}A(\pi_\alpha)f_\alpha(\pi_\alpha)
 \exp\left(\sum\limits_{i\in\alpha}\pi^{(i)}_\alpha
 \chi_{i\alpha}^{(bp)}\right)}
 {\sum\limits_{\pi_\alpha}f_\alpha(\pi_\alpha)
 \exp\left(\sum\limits_{i\in\alpha}\pi^{(i)}_\alpha\chi_{i\alpha}^{(sp)}\right)}.
 \label{pi_av2}
\end{eqnarray}
Using the notation, the ``propagator" and ``vertex" terms can be recast as follows
\begin{eqnarray}
&& P_i=\frac{\prod\limits_{\alpha\ni i}\left[\cosh\zeta_{i\alpha}+
m_i\sinh\zeta_{i\alpha}\right]}
{\left[\cosh\left(\sum\limits_{\alpha\ni
 i}(\chi_{i\alpha}^{(sp)}+\zeta_{i\alpha})/(q_i-1)\right)\right]^{q_i-1}},
 \label{P2}\\
 &&
V_\alpha=\left[\sum\limits_{\pi_\alpha}f_\alpha(\pi_\alpha)\exp
\left(\sum\limits_{i\in\alpha}\pi_\alpha^{(i)}\chi_{i\alpha}^{(sp)}\right)\right]
\Biggl\langle
\prod\limits_{i\in\alpha}\left(1+\frac{(\pi_\alpha^{(i)}-m_i)\tanh\zeta_{i\alpha}}
{1+m_i\tanh\zeta_{i\alpha}}\right) \Biggr\rangle_{\pi_\alpha}.
 \label{V2}
\end{eqnarray}

Of course, the number of terms in the series will grow exponentially with the size, very much like
in the original formulation of the problem. However, wise classification of the terms followed by
selecting (and calculating) a small number of relevant terms allows not only to extract the BP
approximation, but more importantly the leading-order corrections to BP. The leading-order
corrections/terms will be associated with shortest loops on the Tanner graph, and this transparent
geometrical interpretation will be coming through diagrammatic representation of the perturbative
terms. Note, that the diagrammatic technique we develop here is of a special kind. The major
peculiarity of the technique is the non-Gaussian form of the $P$-term in Eq.~(\ref{V1}). Our
approach is technically reminiscent of the celebrated Vaks-Larkin-Pikin approach \cite{66VLP}, used
to calculate non-perturbative corrections to ferromagnetic ground state in magnets.   The reference
is not precise as it only means to emphasize a vague structural relation of our method to the one
introduced in the classical paper \cite{66VLP} where the ``propagator'' $P$-term was also
non-Gaussian.

We are now in a position to discuss a typical structure of the
integrals over $\zeta$ for individual terms (diagrams). We notice
that an individual integral over all possible $\zeta$ variables
always decomposes into a product of independent integrals, each over
the block of variables related to a bit. The simplest integral
corresponds to a bit with all edges connected to it being uncolored:
\begin{eqnarray} &&
I_{0;i}\!\equiv\!\int\prod\limits_{\alpha\ni i}d\chi_{i\alpha} P_i\left(\frac{1+m_i}{2}\right)^{q_i}\exp\left[-\sum\limits_{\alpha\ni
i}\chi_{i\alpha}^{(sp)}\right]\!+\!\left(\frac{1-m_i}{2}\right)^{q_i}
\exp\left[\sum\limits_{\alpha\ni i}\chi_{i\alpha}^{(sp)}\right]
\nonumber\\
&&= \frac{(1-m_i^2)^{(q_i-1)/2}}{2^{q_i-1}}=\frac{
\cosh\left(\sum\limits_{\alpha\ni
i}\chi_{i\alpha}^{(sp)}/(q_i-1)\right)^{1-q_i}}{2^{q_i-1}},
 \label{bit_I0}
\end{eqnarray}
where one uses the saddle-point relation
\begin{eqnarray}
\exp\left(\sum\limits_{\alpha\ni
i}\chi_{i\alpha}^{(sp)}\right)=\left(\frac{1+m_i}{1-m_i}\right)^{(q_i-1)/2}.
\label{sp-rel}
\end{eqnarray}
Combining Eqs.~(\ref{bit_I0},\ref{sp-rel}) and substituting the result in Eq.~(\ref{I1}) we obtain
for the first term in the series for the partition function
\begin{eqnarray}
 Z_0\sim\frac{\prod\limits_\alpha\sum\limits_{\pi_\alpha}f_\alpha(\pi_\alpha)\exp
 \left(\sum\limits_{i\in\alpha}\pi_\alpha^{(i)}\chi_{i\alpha}^{(sp)}\right)}{
 \prod\limits_i
 \cosh\left(\sum\limits_{\alpha\ni
 i}\chi_{i\alpha}^{(sp)}/(q_i-1)\right)^{1-q_i}},
 \label{Z0}
\end{eqnarray}
which exactly reproduces the aforementioned saddle-point result.

For the general $p$-order term one arrives at
\begin{eqnarray} &&
I_{p;i,\{\alpha_l;\ l=1,\cdots, p,\ \alpha_l\in
i\}}\!\equiv\!\int\prod\limits_{\alpha\ni i}d\chi_{i\alpha} P_i
\prod\limits_{l=1}^p\frac{\tanh\zeta_{i\alpha_l}}{\left(1+m_i\tanh\zeta_{i\alpha_l}\right)}
\nonumber\\
&& =\frac{\exp\left[-\sum\limits_{\alpha\ni
i}\chi_{i\alpha}^{(sp)}\right]}{2^{q_i}}
\left(1+m_i\right)^{q_i-p}+(-1)^p
\frac{\exp\left[\sum\limits_{\alpha\ni
i}\chi_{i\alpha}^{(sp)}\right]}{2^{q_i}}
\left(1-m_i\right)^{q_i-p}\nonumber\\
&&
=I_{0;i}\frac{(1-m_i)^{p-1}+(-1)^p(1+m_i)^{p-1}}{2(1-m_i^2)^{p-1}}.
 \label{bit_Ip}
\end{eqnarray}
The resulting expression for the entire series derived directly from
Eqs.~(\ref{I1},\ref{P2},\ref{V2}) and Eq.~(\ref{bit_Ip}) is given by
Eqs.~(\ref{Zseries},\ref{m-fg},\ref{mu-fg}).

\section{Conclusions}
\label{sec:Con}

We conclude with presenting a brief outline for our ongoing and
future research activities on the way of extending the loop calculus
detailed in this paper.

Gauge invariance of the vertex models have been discussed above in
the context of two specific representations we utilized to derive
the loop series formula. However, this important notion allows more
universal and mathematically accurate formulation. For this purpose,
it is convenient to introduce the notion of a graphic tensor and
corresponding graphic trace (convolution). The graphic trace concept
generalizes the standard (in statistical mechanics) transfer matrix
approach to the models defined using arbitrary graphs. This allows
to formulate the loop expansion, the BP equations, as well as the
Bethe free energy, in a gauge-invariant form. The loop expansion
becomes nothing more than a representation of the partition function
as a sum over all possible configurations using a special BP gauge.
The graphical trace and the unifying gauge-invariant approach will
be discussed in details elsewhere \cite{06CCc}. The universal
formulation allows for a natural and straightforward extension of
the loop calculus to more general statistical models that operate
with non-binary alphabets (e.g. Potts model on a graph). This
general problem will also be discussed in \cite{06CCc}.

More generally, we anticipate that the loop calculus can be extended to any classical models
residing in graphs that are formulated in terms of continuous fields of Abelian and non-Abelian
origin, e.g. $O(2)$ and $O(3)$ models, respectively. Moreover, the approach should also work for
Quantum models and fields, e.g. quantum Heisenberg model on a graph. The latter may be of
substantial interest for developing new approaches in quantum information theory.

Loop series offers an exact representation for the partition
function, and also correlation functions, that can be used for
improving approximate algorithms. This should be understood as
follows. Many problems in statistical physics, information and
computer sciences are intractable, in the sense that the number of
steps required to accomplish a computation (i.e. to calculate an
observable) grows exponentially with the system size. Then the issue
of an approximation and related approximate computational algorithm
emerges. Development of a sequence of approximations with gradually
increasing complexity becomes an important task. On the one hand,
the higher is the term in the sequence, the better it approximates
the full answer. On the other hand, the complexity should be linear
or polynomial for at least some number of low-order terms. Given
that the first term in the loop series is the BP term,  which is
known to constitute already a very efficient
approximation/algorithm, one can use the higher-order loop
corrections as a regular way of the BP-algorithm improvement.

The loop series also introduces an explicit BP-measure on the graph: any loop contribution can be
expressed in terms of local objects, magnetizations and irreducable correlation functions
calculated within BP. Therefore, looking for individual loop contributions that dominate the
correction to the bare BP approximation constitutes a particularly attractive and computationally
feasible strategy \cite{06CCS}. As a side remark we note that various options are available for
calculating the bare BP contribution and estimating the magnetizations and irreducible correlation
functions within BP. First of all, one may use the original iterative algorithm of Gallager
\cite{63Gal}. Linear programming approach of \cite{03FWK} is another very attractive possibility at
large signal-to-noise ratios in the cases when the iterative BP does not converge. Finally, one may
develop an iterative relaxation algorithm that is guaranteed to converge to a true minimum of the
Bethe free energy \cite{06SCb}. Such an algorithm is expected to perform better than the linear
programming algorithm at finite temperatures.

We anticipate the proposed scheme to work very well in many cases, especially for the graphs that
are locally tree-like.   The models with long loops emerge naturally in the context of decoding of
LDPC codes \cite{63Gal,99Mac,03RU} and also in $K$-SAT satisfiability problem in computer science
\cite{02MPZ,02MZ,04BZ}.

The loop series can also be very useful for theoretical analysis of
problems with disorder \cite{87MPV}, e.g. of the random graph type
\cite{85VB,01Mon}. The goal here is to calculate the loop
corrections to various disorder-averaged correlation functions. A
particularly interesting question is to differentiate contributions
that originate from loops of different sizes. Depending on the
regime one may expect either dominance of some limited number of
shortest loops, or a distributed effect of many loops. Thus, for the
Viana-Bray model \cite{85VB}, which contains a large number of short
loops, considered in the high-temperature regime, the latter
possibility was reported in the formal $1/N$ replica expansion
\cite{05MR}. The leading correction to the BP expression for the
averaged free energy is dominated by a combined effect of many long
loops. Further analysis of this and other models, especially the
ones corresponding to expurgated ensembles of random graphs modeling
LDPC codes with large girth \cite{02DPTRU}, is required to clarify
the statistical role of loops of different lengths.

Finally, we are optimistic about using the loop calculus developed in this paper for further
analysis and algorithmic exploration of the standard lattice models, i.e regular structures with
many short loops. A particularly interesting yet challenging direction of research would be using
the loop calculus, which naturally differentiates  the loops of different sizes and shapes, for
analysis of the critical point behavior in the lattice models.

We are thankful to M. Stepanov for many fruitful discussions. The
work at LANL was supported by LDRD program, and through start-up
funds at WSU.

\appendix

\section{Bethe free energy}
\label{app:Bethe}

In this Appendix we reproduce a derivation of the Belief Propagation equation based on the Bethe
Free energy variational principle, following closely the description of \cite{05YFW}, e.g.
translating it to our notations. We describe the Bethe free energy approach for the factor graph
model and general vertex models in the two subsequent Subsections, respectively.

\subsection{Bethe Free energy for the factor graph model}
\label{subapp:FG}

In this approach trial probability distributions, called beliefs, are introduced both for bits and
checks $b_i$ and $b_\alpha$, respectively,  where $i=1,\cdots,N$, $\alpha=1,\cdots,M$. Each belief
depends on the corresponding spin realization.  Thus, a belief at a bit actually consists of two
probabilities,  $b_i(+)$ and $b_i(-)$, and we use a natural notation $b_i(\sigma_i)$. There are
$2^k$ beliefs defined at a check, $k$ beeing the number of bits connected to the check, and we
introduce vector noatation ${\bm \sigma}_\alpha=(\sigma_{i_1},\cdots,\sigma_{i_k})$ where
$i_1,\cdots,i_k\in \alpha$ and $\sigma_i=\pm 1$. Beliefs, as corresponding probabilities satisfy
the following inequality constraints
\begin{eqnarray}
 0\leq b_i(\sigma_i),b_\alpha({\bm \sigma}_\alpha)\leq 1,
 \label{ineq}
\end{eqnarray}
the normalization constraints
\begin{eqnarray}
 \sum\limits_{\sigma_i}b_i(\sigma_i)= \sum\limits_{{\bm \sigma}_\alpha} b_\alpha({\bm \sigma}_\alpha)=1,
 \label{norm}
\end{eqnarray}
as well as the consistency (between bits and checks) constraints
\begin{eqnarray}
 \sum\limits_{{\bm \sigma}_\alpha\backslash\sigma_i}b_\alpha({\bm
 \sigma}_\alpha)=b_i(\sigma_i),
 \label{cons}
\end{eqnarray}
where ${\bm \sigma}_\alpha\backslash\sigma_i$ stands for all possible configurations of $\sigma_j$
with $j\in \alpha$, $j\neq i$.

The Bethe Free energy is defined as a difference of the Bethe self-energy and the Bethe entropy,
\begin{eqnarray}
F_{Bethe}=U_{Bethe}-H_{Bethe}, \label{Bethe}
\end{eqnarray}
defined as
\begin{eqnarray}
 && U_{Bethe}=-\sum\limits_\alpha\sum_{{\bm \sigma}_\alpha}b_\alpha({\bm
 \sigma}_\alpha)\ln f_\alpha({\bm \sigma}_\alpha),
 \label{U_Bethe}\\
 && H_{Bethe}=-\sum\limits_\alpha \sum_{{\bm \sigma}_\alpha}b_\alpha({\bm\sigma}_\alpha)
 \ln b_\alpha({\bm\sigma}_\alpha)+\sum\limits_i
 (q_i-1)\sum\limits_{\sigma_i}b_i(\sigma_i)\ln b_i(\sigma_i),
 \label{H_Bethe}
\end{eqnarray}
where ${\bm \sigma}_\alpha=(\sigma_{i_1},\cdots,\sigma_{i_k})$, $i_1,\cdots,i_k\in \alpha$ and
$\sigma_i=\pm 1$. The entropy term for a bit enters Eq.~(\ref{Bethe}) with the coefficient $1-q_i$
to account for the right counting of the number of configurations for a bit: if all entries for a
bit (e.g. into the check term) are counted the total counting should give $+1$ for the bit.

Note, that the definition of $f_\alpha$ according to Eq.~(\ref{U_Bethe}) is not unique. A
convenient choice of the factor function describing an LDPC code would be
\begin{eqnarray}
 f_\alpha({\bm \sigma}_\alpha)\equiv\exp
 \left(\sum\limits_{i\in\alpha} h_i\sigma_i/q_i\right)
 \delta\left(\prod\limits_{i\in\alpha}\sigma_i,1\right).
 \label{factor_LDPC}
\end{eqnarray}

Optimal configurations of beliefs are the ones that minimize the Bethe Free energy (\ref{Bethe})
subject to the constraints (\ref{ineq},\ref{norm},\ref{cons}). Introducing the constraints as the
Lagrange multiplier term to the effective Lagrangian
\begin{eqnarray}
 L=F_{Bethe}+
 \sum\limits_\alpha
 \gamma_\alpha \left(\sum\limits_{{\bm \sigma}_\alpha}
 b_\alpha({\bm \sigma}_\alpha)-1\right)+
 \sum\limits_{i}\gamma_i
 \left(\sum\limits_{\sigma_i}b_i(\sigma_i)-1\right)+
 \sum\limits_i\sum\limits_{\alpha\ni i}\sum\limits_{\sigma_i}
 \lambda_{i\alpha}(\sigma_i)\left(b_i(\sigma_i)-
 \sum\limits_{{\bm \sigma}_\alpha\backslash\sigma_i}b_\alpha({\bm
 \sigma}_\alpha)\right),\label{Lagr}
\end{eqnarray}
and looking for the extremum with respect to all possible beliefs leads to
\begin{eqnarray}
 && \frac{\delta L}{\delta b_a({\bm \sigma}_a)}=0\quad\Rightarrow\quad
 b_\alpha({\bm \sigma}_\alpha)=f_\alpha({\bm\sigma}_\alpha)
 \exp\left[-\gamma_\alpha-1+\sum\limits_{i\in\alpha}\lambda_{i\alpha}(\sigma_i)\right],
 \label{Lba}\\
 && \frac{\delta L}{\delta b_i({\bm \sigma}_i)}=0\quad\Rightarrow\quad
 b_i(\sigma_i)=\exp\left[\frac{1}{q_i-1}\left(\gamma_i+
 \sum\limits_{\alpha\ni i}\lambda_{i\alpha}(\sigma_i)\right)-1\right].
 \label{Lbi}
\end{eqnarray}


Eqs.~(\ref{Lba},\ref{Lbi}) complemented by the normalization and
consistency constraints (\ref{norm},\ref{cons}) form a close system
of BP equations for the $\lambda_{i\alpha}$ variables. In the LDPC
case, described by Eq.~(\ref{factor_LDPC}), BP equations are
traditionally written in terms of $\eta$-fields defined on the edges
according to
\begin{equation}
\eta_{i\alpha}\equiv
\frac{\lambda_{i\alpha}(+)-\lambda_{i\alpha}(-)}{2}+\frac{h_i}{q_i}.
\label{eta_new}
\end{equation}
Substituting Eq.~(\ref{eta_new}) in Eqs.~(\ref{Lba},\ref{Lbi}) one
arrives at the following set of equations for the magnetization at a
bit derived in two different ways
\begin{eqnarray}
&& \sum\limits_{{\bm \sigma}_\alpha}\sigma_i b_\alpha({\bm
\sigma}_\alpha)=\tanh\left(\eta_{i\alpha}+\tanh^{-1}\left(\prod\limits_{j\in\alpha}^{j\neq
i}\tanh\eta_{j\alpha}\right)\right), \label{mag1}\\
&& \sum\limits_{\sigma_i}\sigma_i
b_i(\sigma_i)=\tanh\left(\frac{\sum_{\alpha\ni
i}(\lambda_{i\alpha}(+)-\lambda_{i\alpha}(-))}{2(q_i-1)}\right).
\label{mag2}
\end{eqnarray}
Equating the right hand sides of Eqs.~(\ref{mag1},\ref{mag2}), using
Eq.~(\ref{eta_new}) and making some simple manipulations  one
derives
\begin{eqnarray}
 \eta_{i\alpha}=h_i+\sum\limits_{\beta\ni i}^{\beta\neq \alpha}
 \tanh^{-1}\left[\prod\limits_{j\in\beta}^{j\neq
 i}\tanh\eta_{j\beta}\right],
 \label{bp1new}
\end{eqnarray}
that is the BP system of equations for LDPC codes written in its
standard form. (These equations are often described in the coding
theory literature as stationary point equations for the sum product
algorithm.)

\subsection{Bethe free energy for the general vertex model}
\label{subapp:GVG}

The variational approximation for the model that generalizes the factor graph case discussed in
Appendix \ref{subapp:FG}, reads as follows. One minimizes the following Bethe free energy
\begin{eqnarray}
 F_{gvm}=\sum_a \sum_{{\bm \sigma}_a}b_a({\bm \sigma}_a)
 \ln\left(\frac{b_a({\bm \sigma}_a)}{f_a({\bm \sigma}_a)}\right)-\sum_{a,c}^{c\in
 a}\sum_{\sigma_{ac}}
 b_{ac}(\sigma_{ac})
 \ln b_{ac}(\sigma_{ac}),
 \label{Fgvm}
\end{eqnarray}
with respect to $b_a({\bm \sigma}_a), b_{ac}(\sigma_{ac})$ fields under the conditions
\begin{eqnarray}
 && \forall\ a,c;\ c\in a:\ \ 0\leq b_a({\bm \sigma}_a),b_{ac}(\sigma_{a,c}\leq 1,\label{gvm_in}\\
 && \forall\ a,c;\ c\in a:\ \ \sum_{{\bm \sigma}_a} b_a({\bm \sigma}_a)=
 \sum_{\sigma_{a,c}} b_{ac}(\sigma_{a,c})=
 1,\label{gvm_1}\\
 && \forall\ a,c;\ c\in a:\ \ b_{ac}(\sigma_{ac})=\sum_{{\bm \sigma}_a\setminus\sigma_{ac}} b_a({\bm \sigma}_a)=
 \sum_{\sigma_c\setminus\sigma_{ac}} b_c({\bm \sigma}_c),\label{gvm_2}
\end{eqnarray}
where as usual we assume $\sigma_{ac}=\sigma_{ca}$. The second term on the rhs of Eq.~(\ref{Fgvm})
is the entropy term which takes care of the ``double" counting" of the link contribution: any link
enters twice in the entropy part of the first term on the rhs of Eq.~(\ref{Fgvm}).

Extension of these formulas to the orientable vertex model case is straightforward. It is achieved
by partitioning the entire family of vertices $\{a\}$ into two sub-families $\{i\}$ and
$\{\alpha\}$. After that one just needs to replicate Eqs.~(\ref{Fgvm},\ref{gvm_in},\ref{gvm_1}) in
the bit and check versions respectively, while Eq.~(\ref{gvm_2}) adopts the following form
\begin{eqnarray}
\forall\ i,\alpha;\ i\in \alpha:\ \ \sum_{{\bm
\sigma}_i\setminus\sigma_{i\alpha}} b_i({\bm \sigma}_i)=
 \sum_{{\bm \sigma}_\alpha\setminus\sigma_{i\alpha}} b_\alpha({\bm \sigma}_\alpha).\label{gvm_2}
\end{eqnarray}

Furthermore, considering the case of the orientable vertex model and substituting a particular form
of the $f_i({\bm \sigma}_i)$ correspondent to Eq.~(\ref{fg-ov}), we finds that Eq.~(\ref{Fgvm})
turns into Eq.~(\ref{Bethe},\ref{U_Bethe},\ref{H_Bethe}) under a natural substitution
\begin{eqnarray}
b_i({\bm \sigma}_i)=\left\{
\begin{array}{cc} b_i(\sigma_i), & \sigma_{i\alpha}=\sigma_{i\beta}\ \ \forall \alpha,\beta\ni i\ \
\\
0, & \mbox{  otherwise.}\end{array}\right. \label{b_i}
\end{eqnarray}

\section{Single loop example}
\label{app:single}

\begin{figure} [b]
\includegraphics[width=0.45\textwidth]{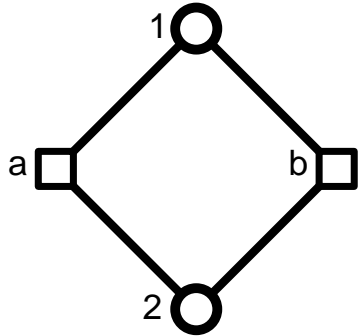}
\caption{Factor graph for the single loop model consisting of 2
nodes and 2 bits.} \label{fig:2_2}
\end{figure}

This Appendix serves an illustrative purpose: We show directly how
the loop formula (\ref{Zseries}) works for a simple example of the
factor graph model (\ref{p1}) with a single loop (two bits and two
checks, see Fig.~\ref{fig:2_2}). For this simple model the
belief-propagation equations (\ref{Lba},\ref{Lbi}) adopt the
following form
\begin{eqnarray}
 && \mbox{for}\ \ \alpha,\beta=a,b; \ \ \beta\neq\alpha:\ \ \
 b_\alpha(\sigma_1,\sigma_2)=\frac{f_\alpha(\sigma_1,\sigma_2)
 d_{1\beta}^{\sigma_1/2}
 d_{2\beta}^{\sigma_2/2}}
 {\sum_{\sigma'_{1,2}}
 f_\alpha(\sigma'_1,\sigma'_2)
 d_{1\beta}^{\sigma'_1/2}
 d_{2\beta}^{\sigma'_2/2}},
 \label{eq2_1}\\
 &&
 \mbox{for}\ \ i=1,2:\ \ \
 b_i(\sigma_i)=\frac{
 d_{i a}^{\sigma_i/2}
 d_{i b}^{\sigma_i/2}}
 {\sum_{\sigma'_i}
 d_{i a}^{\sigma'_i/2}
 d_{i b}^{\sigma'_i/2}}
 \label{eq2_2}
\end{eqnarray}
where the factor functions $f_a(\sigma_1,\sigma_2)$,
$f_b(\sigma_1,\sigma_2)$, defined for $\sigma_{1,2}=\pm 1$ are
considered to be arbitrary. Eqs.~(\ref{eq2_1},\ref{eq2_2})
complemented by the consistency conditions (\ref{cons}) are reduced
to the set of quadratic equations that can be solved explicitly
yielding
\begin{eqnarray}
 &&     d_{1a}=\Biggl(-f_a(-,-)f_b(-,-)-f_a(-,+)f_b(-,+)+f_a(+,-)f_b(+,-)+f_a(+,+)f_b(+,+) \nonumber\\
 &&     \Biggl[4\left(f_b(-,-)f_a(-,+)+f_b(+,-)f_a(+,+)\right)\left(f_a(-,-)f_b(-,+)+f_a(+,-)f_b(+,+)\right) \nonumber\\
 &&     (f_a(-,-)f_b(-,-)-f_a(-,+)f_b(-,+)+f_a(+,-)f_b(+,-)-f_a(+,+)f_b(+,+))^2\Biggr]^{1/2}\Biggr)\nonumber\\
 &&     \times\left[2(f_a(-,-) f_b(+,-)+ f_a(-,+) f_b(+,+))\right]^{-1},
 \label{d1a}
\end{eqnarray}
where the BP expressions for $d_{1b},d_{2a}$ and $d_{2b}$ can be derived making proper permutations
of indices and arguments in Eq.~(\ref{d1a}). Using these solutions we arrive at the following
expressions for the partition function calculated within the BP-approach:
\begin{eqnarray}
 &&
 Z_0=\frac{\prod_\alpha\sum_{\sigma_{1,2}}^{\beta\neq\alpha}
 f_\alpha(\sigma_1,\sigma_2)d_{1\beta}^{\sigma_1/2}d_{2\beta}^{\sigma_2/2}}
 {\prod_i\sum_{\sigma_i}d_{ia}^{\sigma_i/2}d_{ib}^{\sigma_i/2}}\nonumber\\ &&
 =\frac{1}{2}\Biggl(f_a(-,-)f_b(-,-)+f_a(-,+)f_b(-,+)+f_a(+,-)f_b(+,-)+f_a(+,+)f_b(+,+)
 \nonumber\\ &&
 +\Biggl[4(f_b(-,-)f_a(-,+)+f_b(+,-)f_a(+,+))(f_a(-,-)f_b(-,+)+f_a(+,-)f_b(+,+))
 \nonumber\\ &&
 +(f_a(-,-)f_b(-,-)-f_a(-,+)f_b(-,+)+f_a(+,-)f_b(+,-)-f_a(+,+)f_b(+,+))^2\Biggr]^{1/2}\Biggr).
 \label{Z02_2}
\end{eqnarray}
Bit magnetizations as well as irreducible correlation functions at the checks are found upon direct
substitution of (\ref{d1a}) and similar expressions for the other $d$-variables in terms of the
factor-functions into
\begin{eqnarray}
 && i=1,2:\ \ \ m_i=\sum_{\sigma_i}\sigma_i b_i(\sigma_i),
 \label{m12}\\
 && \alpha=a,b:\ \ \ \mu_\alpha=\sum_{\sigma_1,\sigma_2}
 (\sigma_1-m_1)(\sigma_2-m_2)b_\alpha(\sigma_1,\sigma_2).
 \label{muab}
\end{eqnarray}
Substituting these results, together with Eq.~(\ref{Z02_2}), into the loop expression
Eq.~(\ref{Zseries}) for the model partition function we obtain
\begin{eqnarray}
 && Z=Z_0\left(1+\frac{\mu_a\mu_b}{(1-m_1^2)(1-m_2^2)}\right)\nonumber\\
 &&=f_a(-,-)f_b(-,-)+f_a(-,+)f_b(-,+)+f_a(+,-)f_b(+,-)+f_a(+,+)f_b(+,+),
 \label{Z2_2}
\end{eqnarray}
which coincides with the exact expression for the model partition function that can be evaluated
directly.


\begin{thebibliography}{99}
\bibitem{Models} Examples of inference problems of information and computer sciences expressed in graphical
terms can be found in \cite{01KFL,01For,04Loe}. Notice, however, that the simple modeling
terminology we define and use in the manuscript is a bit different from the one used in the
information science. Thus, our general vertex model is equivalent to what is called ``normal"
factor graph model in \cite{01For} and ``Forney's factor graph" model in \cite{04Loe}. Our
orientable graph model is equivalent to a normal factor graph which has been obtained by taking a
standard bipartite factor graph of \cite{01KFL}, and turning all the variable nodes into the
so-called equality factor nodes \cite{01For}.
\end{thebibliography}
\end{document}